%% file: main.tex
\documentclass[runningheads]{llncs}
\usepackage{amsmath}
\usepackage[T1]{fontenc}
\usepackage{graphicx}
\usepackage{amssymb}  
\usepackage{array}
\usepackage{hyperref} 
\usepackage{appendix}

\usepackage{multicol}
\usepackage{multirow}
\usepackage[table]{xcolor}
\definecolor{lightGray}{gray}{0.9}   
\definecolor{mediumGray}{gray}{0.8}   
\makeatletter
\newcommand{\equalcontribmark}{\textsuperscript{*}}
\makeatother

\let\oldpm\pm
\renewcommand{\pm}{%
  \mathbin{%
    \mathchoice
      {\raisebox{0.3ex}{\scalebox{0.5}{\ensuremath{\oldpm}}}}%
      {\raisebox{0.3ex}{\scalebox{0.5}{\ensuremath{\oldpm}}}}%
      {\raisebox{0.2ex}{\scalebox{0.5}{\ensuremath{\oldpm}}}}%
      {\raisebox{0.2ex}{\scalebox{0.5}{\ensuremath{\oldpm}}}}%
  }%
}
\newcommand{\smallscript}[1]{\scalebox{0.8}{\ensuremath{\scriptstyle #1}}}
\begin{document}
\title{GrInAdapt: Scaling Retinal Vessel Structural Map Segmentation Through Grounding, Integrating and Adapting Multi-device, Multi-site, and Multi-modal Fundus Domains}

%
%
\author{  Zixuan Liu\inst{1}\equalcontribmark \and
  Aaron Honjaya\inst{1}\equalcontribmark \and
Yuekai Xu\inst{1}\equalcontribmark \and Yi Zhang\inst{2} \and Hefu
Pan\inst{2} \and Xin Wang\inst{3} \and Linda G. Shapiro\inst{1} \and Sheng Wang\inst{1} \and Ruikang K. Wang\inst{2,4} 
}
\authorrunning{Zixuan Liu et al.}
\institute{Paul G. Allen School of Computer Science and Engineering, University of Wasghington, Seattle, WA, 98195, USA \\\email{\{zucksliu,swang\}@cs.washington.edu} \and
Department of Bioengineering, University of Washington \and
Department of Electrical and Computer Engineering, University of Washington \and Department of Ophthalmology, University of Washington\\ \email{wangrk@uw.edu}}
%


\maketitle              
%
\begingroup
\renewcommand\thefootnote{*}%
\footnotetext{Equal contribution.}%
\endgroup

\begin{abstract}
\input{abstract}
\keywords{Domain adaptation  \and Retinal imaging \and Vessel segmentation.}

\end{abstract}
\section{Introduction}
\begin{figure}[ht]
    \centering
    \setlength{\abovecaptionskip}{-3pt}  
    \setlength{\belowcaptionskip}{-3pt}    
    \includegraphics[width=\linewidth]{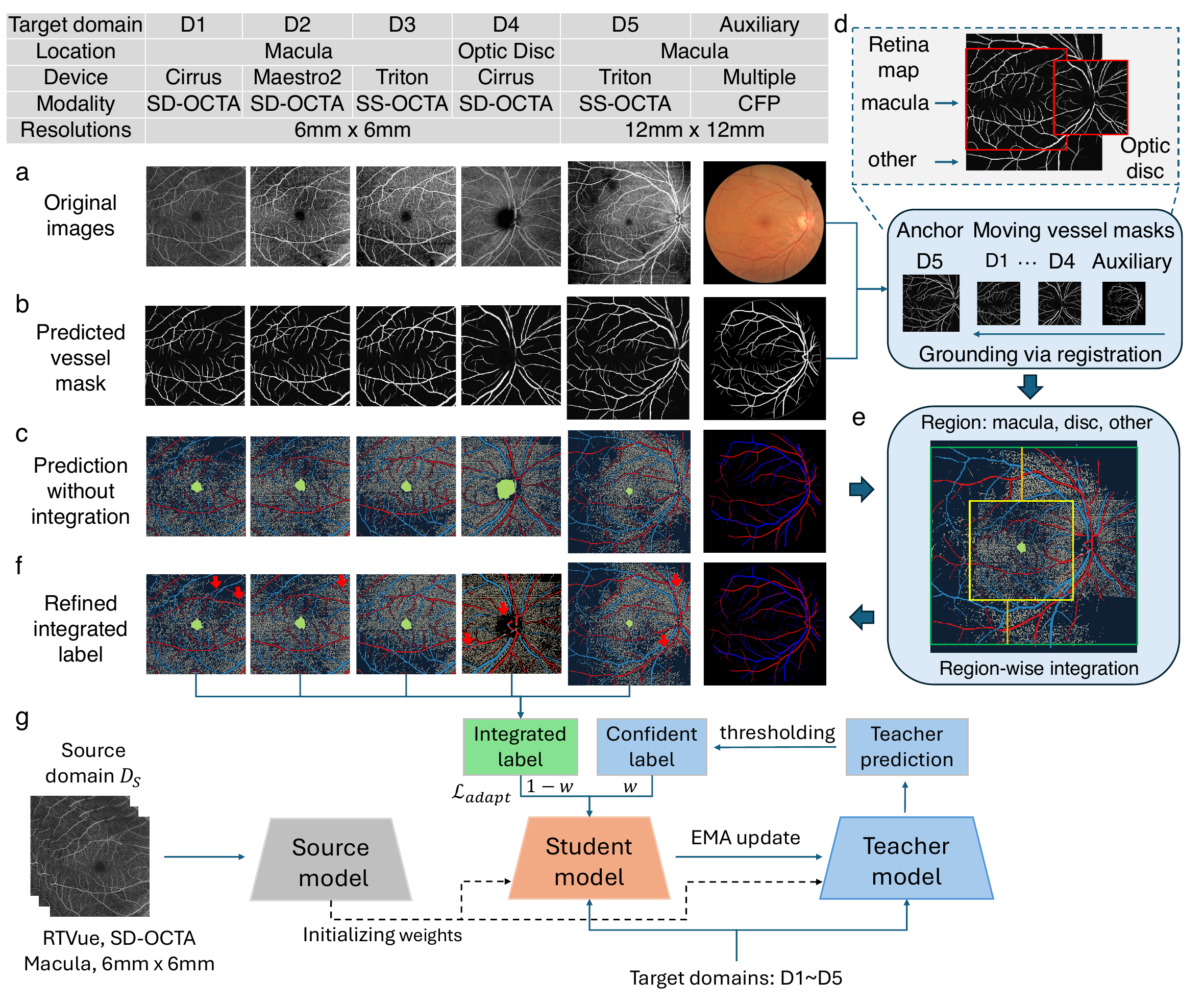}
    \caption{Overview of GrInAdapt. \textbf{a}. Paired \textit{en face} images from different domains. \textbf{b}. Predicted vessel masks used for registration. \textbf{c}. Prediction from souce models used for integration. \textbf{d}. Three regions on retina - macula, optic disc and other. \textbf{e}. Different predictions are integrated based on the region. \textbf{f}. Refined integrated labels for adapatation. \textbf{g}. Adaptation process with student-teacher and adaptive label merging.}  
    \label{fig:fig1}
\end{figure}
\input{introduction_v2}

\input{Prelim_v2}

\section{GrInAdapt: Grounding, Integrating and Adapting multiple domains}
\input{Method_v2}

\section{Experiments}
\input{experiments}

\section{Conclusion}
\input{conclusion}

%
%

\bibliographystyle{splncs04}
\bibliography{mybibliography}

\newpage
\input{Supp/Supp_result}
\input{Supp/Supp_mat}

\input{Supp/Supp_reg}
\input{Supp/Supp_integrate}
\input{Supp/Supp_adapt}

\input{Supp/Supp_other}

\end{document}

%% file: abstract.tex
Retinal vessel segmentation is critical for diagnosing ocular conditions, yet current deep learning methods are limited by modality-specific challenges and significant distribution shifts across imaging devices, resolutions, and anatomical regions. In this paper, we propose GrInAdapt, a novel framework for source-free multi-target domain adaptation that leverages multi-view images to refine segmentation labels and enhance model generalizability for optical coherence tomography angiography (OCTA) of the fundus of the eye. GrInAdapt follows an intuitive three-step approach: (i) grounding images to a common anchor space via registration, (ii) integrating predictions from multiple views to achieve improved label consensus, and (iii) adapting the source model to diverse target domains. Furthermore, GrInAdapt is flexible enough to incorporate auxiliary modalities—such as color fundus photography—to provide complementary cues for robust vessel segmentation. Extensive experiments on a multi-device, multi-site, and multi-modal retinal dataset demonstrate that GrInAdapt significantly outperforms existing domain adaptation methods, achieving higher segmentation accuracy and robustness across multiple domains. These results highlight the potential of GrInAdapt to advance automated retinal vessel analysis and
support robust clinical decision-making. 

%% file: introduction_v2.tex
Retinal vessel segmentation \cite{fraz2012blood} and quantification \cite{lee2018quantification} are critical for early diagnosis and management of ocular diseases such as retinal vascular changes \cite{cardoso2016systematic} and diabetic retinopathy \cite{tey2019optical}. Accurate delineation of arteries and veins \cite{li2024octa} facilitates disease detection and monitoring of disease progression \cite{cunha2021central}. Various imaging modalities—including color fundus photography (CFP) \cite{zhou2021learning}, and optical coherence tomography angiography (OCTA) \cite{li2024octa}—offer complementary views of the retinal vasculature \cite{chen2021retinal}. However, each modality has inherent trade-offs: CFP provides a wide field-of-view but often misses fine capillary details, while OCTA delivers high-resolution, three-dimensional information at the cost of being more sensitive to noise and artifacts.

Despite advances in deep learning for segmenting major vessels from CFP \cite{zhou2021learning,shi2024one} and extracting fine-grained capillary structures from OCTA \cite{gao2022deep}, current methods remain limited by two factors. First, single-modality approaches are constrained by modality-specific shortcomings—for example, CFP’s limited sensitivity to capillaries and OCTA’s vulnerability to imaging artifacts \cite{enders2019quantity,hormel2021artifacts}. Second, models trained on a single source domain struggle to generalize to new devices, fields-of-view, and clinical settings, leading to performance degradation due to distribution shifts \cite{farahani2021brief,kouw2019review}. While prior work has explored domain adaptation \cite{tang2023source,chen2021source,huai2023context} and multi-modal integration \cite{li2015cross,zhao2024global,liu2024octcube}, these approaches typically address non-vessel structure, low-resource, or single-domain scenarios \cite{ai2024ai,Owsleye097449}.

In this work, we go one step forward to explore a higher-resource but more ambitious setting: how to widely generalize a source model to multiple target domains through a dataset with multi-view images from each target domain. We propose a powerful framework, GrInAdapt, with the key idea to leverage multi-view images to obtain better labels and refine the model across multiple target domains. Specifically, GrInAdapt follows an intuitive three-step approach: first, all images are grounded to an anchor space through registration; second, the labels are refined by integrating predictions from multiple views; and finally, the model is adapted to the target domains. GrInAdapt is simple and robust—even a basic registration  applied to imperfect vessel mask predictions can be effective. It demonstrates superiority in achieving label consensus and reducing noise caused by imaging artifacts and low imaging quality. Moreover, GrInAdapt is flexible: it 
can be easily equipped with ensemble learning and extended to a multi-modal, multi-view setting to further reinforce label robustness via auxiliary modalities.

Extensive experiments on the AI-READI dataset \cite{ai2024ai} with multi-device, multi-site, and multi-modal retinal images demonstrate that GrInAdapt consistently improves the source model by on average a 4\% Dice score increase and a 0.42 ASSD reduction Ablation studies confirm that each component of our three-step approach—registration, label integration, and adaptation—are robust and contributes significantly to the observed performance improvements. Further evaluations reveal robust generalization across sites, unseen locations, and different resolutions, with quantitative gains of 4\% and 4.6\% in Dice score. These findings underscore GrInAdapt’s potential to enhance automated retinal analysis.

%% file: Prelim_v2.tex
\section{Preliminaries of Multi-target Domain Adaptation}

Although the setting described here applies to various medical imaging modalities, in this work we focus on fundus OCTA and CAVF segmentation - Capillaries, Artery, Vein and Foveal avascular zone (FAZ) \cite{li2024octa}. Let 
\(
\mathcal{S} = \{(x^s_i, y^s_i)\}_{i=1}^{N_s}
\)
denote the C-class segmentation labeled source domain, and let there be $M$ unlabeled target domains 
\(\mathbf{T}^M = \{\mathcal{T}_j\}_{j=1}^{M},\) where each \(\mathcal{T}_j\) is defined as \(\mathcal{T}_j = \{x^{t_j}_i\}_{i=1}^{N_{t_j}}\).
In the source-free adaptation setting, only the pre-trained source model \(f_{\theta_{S}}\) is available during adaptation. The objective of multi-target domain adaptation (MDA) is to adapt \(f_{\theta_{S}}: \mathcal{X}\rightarrow \mathcal{Y} \) so that it performs well across all target domains $\{\mathcal{T}_j\}^{M}_{j=1}$. Adapting to multiple target domains is more challenging because the distribution shift is larger and more diverse. For example, in OCTA imaging, when taking macula-centered \(6 \times 6\) scans from certain device as the source domain, target domain shifts can arise from several factors (Fig. \ref{fig:fig1}a): imaging quality shift (different devices, D1-D3), location shift (D4), resolution shift (D5), and possible shift caused by the variance of clinicians' skill across different sites.   

\noindent\textbf{Subject-level multi-target image pair assumption.} Previous DA methods usually focus on a single domain. A naive baseline for MDA is thus to simply treat \(\mathbf{T}^M\) as a larger single target domain \cite{farahani2021brief,kouw2019review}. However, downgrading adaptation performance is expected as the number and diversity of target domains increase. We thus relax data availability, allowing subject-level data pairs across multiple domains, e.g., an eye being imaged from multiple devices, resolutions, and locations. Specifically, given a subject set $\{s_k\}^{V}_{k=1}$, for each subject \(s_k\), an image is acquired in every target domain \(j\), i.e.,
\(
\{x^{t_j}_{s_k}\}_{j=1}^{M}, x^{t_j}_{s_k} \in \mathcal{T}_j.
\)
This cross-target paired information is essential for aligning representations across domains and mitigating the effects of the aforementioned shifts.
Optionally, paired data from auxiliary modalities $\mathcal{A}$ such as color fundus photography (CFP) can also be incorporated (D6, Fig. \ref{fig:fig1}a). 
We assume that $\forall s_k \in V$, paired data $z_{s_k} \in \mathcal{A}$ and a source-free model $f_{\theta_A}: \mathcal{Z}\rightarrow \mathcal{W}$ that can generate a $C_W$-class auxiliary segmentation label $w\in \mathcal{W}$ are available.
Here, the primary and auxiliary label space \(\mathcal{Y}\) and \(\mathcal{W}\)
share common semantic classes, i.e., \(\mathcal{Y} \cap \mathcal{W} \supseteq \{\text{Artery, Vein}\}\), allowing the auxiliary information to provide complementary cues for predicting the primary segmentation label \(y \in \mathcal{Y}\).

%% file: Method_v2.tex
In this section, we introduce GrInAdapt, our framework designed to robustly generalize a pre-trained source model to multiple target domains. GrInAdapt is a three-step process: grounding, integrating, and adapting, each addressing specific challenges arising from domain shifts in retinal vessel segmentation.

\subsection{Grounding via Segmented Mask Registration}
Inspired by \cite{ding2020weakly,wang2023bingo}, we proposed a grounding step that aligns target images of the same subject to a common anchor space to mitigate spatial variability arising from heterogeneous acquisition protocols in multi-target domain adaptation. Given a data bag \(\mathcal{I}_{s_k}=\{I_j\}^{M+1}_{j=1}=\{x^{t_j}_{s_k}\}^{M}_{j=1} \cup \{z_{s_k}\}\) of subject $s_k$, we assumed for every two images $(I_{anc}, I_{mov})$, there exists a spatial transform $h \in \mathcal{H}$ to map the coordinate system of $I_{mov}$ to that of $I_{anc}$. In our framework, we first picked an anchor image $I_{anc} \in \mathcal{I}_{s_k}$ and a registration function $R(g): g(\mathcal{X}_{anc}) \times g(\mathcal{X}_{mov}) \rightarrow \mathcal{H}$ to estimate the spatial transformation $h \in \mathcal{H}$ that aligns the coordinates
of all other moving images $I_{mov} \in \mathcal{I}_{s_k} \backslash \{I_{anc}\}$ to the coordinates of the anchor image. 

We only require an invertible transformation $h_j$: $\forall h_j \in \mathcal{H}$, there exists an accessible inverse transform $h_j^{-1}$ to map the registered moving image back for the adapting process later. 
To ensure registration quality, we used the predicted binary vessel probability maps $g(x^{t_j}_{s_k})=\hat{p}^{t_j, ves}_{s_k}$
as they provide smooth morphological predictions (Fig. \ref{fig:fig1}b).
The resulting $h_j$ and $h_j^{-1}$ can be used to register the actual segmentation probability map $\hat{p}_{s_k}^{t_j} \in \mathbb{P}^{C}_{\mathcal{X}}$ (Fig. \ref{fig:fig1}c). A similar process can also be applied to $\hat{p}_{s_k}^{z} \in \mathbb{P}^{C_W}_{\mathcal{X}}$. By doing so, the grounded mask predictions from multiple domains can be efficiently aggregated to share learned information. 

\subsection{Integrating via Reliable Region-wise Label Merging}
Let \(\tilde{p}^{t_j}_{s_k}=h(\hat{p}_{s_k}^{t_j})\) and \(\tilde{p}^{a_j}_{s_k}=h(\hat{p}_{s_k}^{a_j})\) denote the registered segmentation probability maps for subject \(s_k\) from the target and auxiliary modalities, respectively. With all predictions aligned in a common anchor space, our objective is to refine the segmentation output by fusing these diverse predictions in a manner that is sensitive to regional variations in reliability.
The key idea is very intuitive: only reliable domains will be considered for integration. Let \(L\subset \mathcal{X}\) denote a defined region within the anchor-space image (e.g., the macular region, optic disc region, or remaining regions), and let \(J_L \subset \{1,\dots, M+M_A\}\) be the index set corresponding to the selected images that have reliable contribution to region \(L\). 
For each pixel \(u \in L\), the refined soft probability map is computed by averaging the softmax probabilities from the selected predictions (Fig. \ref{fig:fig1}d-f):
\begin{equation}
\hat{p}^{\text{soft}}_{s_k}(u) = h^{-1}(\frac{1}{|J_L|} \sum_{j \in J_L} \tilde{p}^{j}_{s_k}(u)),\quad \hat{y}^{\text{hard}}_{s_k}=\arg\max_{c\in\{1,\dots,C\}}\hat{p}^{\text{soft},c}_{s_k}(u)
\label{eqn:merge}
\end{equation}
where \(\tilde{p}^{j}_{s_k}(x) \in \mathbb{P}^C\) is the probability vector (over \(C\) classes) predicted by the \(j\)-th image for subject \(s_k\). Note that ensemble learning technique such as bagging \cite{sewell2008ensemble} multiple model replica by averaging their predictions can be easily plugged in. The outputs \(\hat{p}^{\text{soft}}_{s_k}(x)\) and \(\hat{y}^{\text{hard}}_{s_k}(x)\) thereby serve as robust pseudo-labels transformed back to the original space for subsequent adaptation. 


Choosing $L$ and corresponding $J_L$ is crucial. We split the retina map into 3 regions: macula, optic disc and other area (Fig. \ref{fig:fig1}d,e). In the macular region, \(\mathcal{J}_L\) includes only macula-centered OCTA predictions while excluding CFP images because they lack the fine vessel details for accurate segmentation in the central macula. Conversely, in the optic disc region, we relied on optic disc–centered OCTA predictions in combination with CFP images, which have been demonstrated to yield precise vessel delineations in this area. For all other regions, \(\mathcal{J}_L\) comprises both CFP and OCTA predictions to benefit from both modalities.

\subsection{Adapting via Teacher-Student Learning with Integrated Label}
Unlike conventional domain adaptation methods that primarily focus on denoising the source model's predictions, our approach leveraged the integrated labels to correct and improve the model (Fig. \ref{fig:fig1}g). In the adaptation stage, the integrated label \(\hat{y}^{\text{hard}}_{s_k}\) is used for pseudo-supervision. Specifically, initiating $f_\theta$ with the source model $f_{\theta_S}$, for each pixel $u \in \mathcal{X}$, we defined a segmentation loss as:
\begin{equation}
\mathcal{L}_{seg} = \frac{1}{|\mathcal{X}|}\sum_{u \in \mathcal{X}}\ell(f_\theta(u), \hat{y}^{\text{hard}}_{s_k, u}), \quad \ell(f(u), y)= \text{Dice}(f(u), y) + \lambda\text{CE}(f(u),y)
\label{eqn:segloss}
\end{equation}

However, as \(\hat{y}^{\text{hard}}_{s_k}\) may still contain noise, we further incorporated a confidence loss exploiting high-confidence model predictions \cite{chen2021source} thresholded by $\tau_c$:
\begin{equation}
\mathcal{L}_{conf} = \frac{1}{|\mathcal{X}|}\sum_{x \in \mathcal{X}} \mathbb{I}\Bigl(\max_{c} f^c_{\theta}(u) > \tau_c\Bigr) \cdot \ell\Bigl(f_{\theta}(u), \hat{y}^{\text{hard}}_{s_k, u}\Bigr),
\label{eqn:confloss}
\end{equation}

To prevent the model from reinforcing only its own potentially erroneous predictions, we employed a teacher-student framework following \cite{french2017self,tang2023source}. The teacher model \(f_{\theta_T}\) provides more stable targets using weakly-augmented samples and is updated through an exponential moving average (EMA) of the student parameters, which train on strongly-augmented samples: $\theta_T \leftarrow \alpha \theta_T + (1-\alpha)\theta$,
with \(\alpha\) as the smoothing coefficient. The overall adaptation loss is then formulated as:
\begin{equation}
\mathcal{L}_{adapt} = \lambda(t)\, \mathcal{L}_{conf} + \bigl(1-\lambda(t)\bigr)\, \mathcal{L}_{seg},
\label{eqn:adaptloss}
\end{equation}
where \(\lambda(t)\) denotes a cosine annealing weight schedule over \(E\) epochs. 
\(\lambda(t)\) starts with a low value to prioritize learning from the integrated labels and gradually increases to place greater emphasis on the teacher’s high-confidence predictions.

%% file: experiments.tex
\subsection{Experimental Setup for Multi-target Domain Adaptation}

\input{dice_assd_table_results/dice_no_site_avf}
\noindent\textbf{Source and Multiple Target Data Domains.} 
We aimed to segment out a 2D \textit{en face} CAVF mask, while capillaries serve as an auxiliary task. To evaluate GrInAdapt, we picked the OCTA-500 \cite{li2024octa} dataset as our source domain and 5 modalities of OCTA scans from the AI-READI dataset \cite{ai2024ai,Owsleye097449} as our multi-target domains. The OCTA-500 dataset contains 300 6mm$\times$6mm macula-centered OCTA scans captured by RTVue SD-OCTA device, each paired with a CAVF annotation. The AI-READI dataset has 1,060 patients, 2,112 eyes and in total 10,496 images from 3 sites (UW, UAB, UCSD) with 3 imaging modalities of 6mm$\times$6mm macula-centered OCTA (imaged by Maestro2 SD-OCTA, Cirrus SD-OCTA, Triton SS-OCTA), one 6mm$\times$6mm Optic disc centered Cirrus SD-OCTA, and one 12mm$\times$12mm macula-centered Triton SS-OCTA modality with no CAVF annotations. 
3D OCTA flow, 3D OCT structural volume, and 2D superficial \textit{en face} projection map are provided for all domains from both datasets, and the AI-READI dataset has at least one paired 2D \textit{en face} CFP image for each scan. To facilitate the evaluation process, we selected 16 patients from three sites with balanced lateral distribution based on imaging quality, and annotated the CAVF masks by professionally trained ophthalmologists, resulting in a test set of 80 images held-out throughout the training process.

\noindent\textbf{Model Architecture.} 
For the source segmentation model $f_{\theta_S}$, taking both 3D OCTA, 3D OCT and, 2D projection map as input, it is a 4-block, 3D Res-UNet equipped with IPN v2 architecture~\cite{ronneberger2015u,li2024octa} and enhanced with a 2D branch that processes superficial retinal features. The 3D branch is designed to capture volumetric contextual information from OCTA and OCT data, while the 2D branch extracts complementary features from en face projections. 

\noindent\textbf{Implementation Details.} We first trained $f_{\theta_S}$ using the OCTA-500 dataset with a 240:10:50 train:val:test split via the standard segmentation loss $l(f(x),y)$ in Eqn. (\ref{eqn:segloss}), and utilized a trained $f_{\theta_A}$ from \cite{zhou2021learning}. $f_{\theta_S}$ and $f_{\theta_{A}}$ were used to predict artery-vein masks and generate binary vessel masks for OCTA scans and CFP images. We used a simple key point-based registration with affine transform and AKAZE detector\cite{alcantarilla2011fast,shi2024one}. We randomly chose an anchor image and monitored if an image passed registration by thresholding the translation, scaling, shear and perspective factors extracted from the transform. 
For each subject, we iterated the anchor image until it successfully registered all images, or failed on all possible anchors. We only picked subjects with all images successfully registered for label integration. For the integration, we used images with a smaller field of view - 6mm$\times$6mm macula-centered and optic disc centered images to split the region. To improve label robustness, we trained 3 replicas of $f_{\theta_S}$ and ensembled them by taking the average prediction for each registered image. We only transformed the artery and vein classes back to update the original prediction as we found FAZ labels tended to shrink to the intersection across different domains. 
In the adaptation stage, we set $\alpha$=0.995, a learning rate of 8$\times $$10^{-5}$, and the $\lambda(t)$ schedule to be from 0.1 to 0.9 over 3 epochs. We let $\lambda$=1, $\tau_{\text{artery}}$=$\tau_{\text{vein}}$=$0.5$, $\tau_{\text{capillaries}}$=0, and a region-wise FAZ thresholding scheme for $\tau_{\text{faz}}$. Different levels of Gaussian noise were added to images as weak and strong augmentations.

\subsection{Experimental Results}
\noindent\textbf{Registration results.} GrInAdapt has a high success rate of registration even with a simple, automated, and parameter-free key-point based registration algorithm. Of 2,112 eyes, 1,562 from 868 patients successfully has all of their images registered, resulting in 7,303 registered images with a 74.2\% subject-level and 81,8\% image-level all-success rate. Among the failure image cases, more than 60\% of them were found to have too poor imaging quality to generate a valid vessel mask through quality check. This validated the robustness of the design and provided us a large cohort for label integration and adaptation.
\input{dice_assd_table_results/both_only_site_assd_all_1deci}

\noindent\textbf{Domain adaption performance.} We compared GrInAdapt with 5 baselines to examine its effectiveness (Table \ref{tab:dce_nosite_avf}). We first validated the good quality of integrated label (D1-D3, D5) especially on Artery and Vein, since it facilitates adaptation for most domains by serving as the target label. By utilizing it, GrInAdapt enhanced the source model with an average of 4\% improvement of Dice score. The improvement scale is generally consistent among the artery and vein classes across the macular region of different domains (D1-D3), suggesting its ability to simultaneously fit multiple domains. Notably, the 12mm$\times$12mm field (D5) had a larger 5.3\% average improvement, indicating the power of integrating vessel details from the smaller 6mm$\times$6mm view to the wider but less detailed 12mm$\times$12mm view. For the challenging optic disc domain (D4), GrInAdapt successfully learned to identify the area and not predict FAZ. We further witnessed a substantial improvement on artery and vein performance, especially when integrated label performances are largely downgraded and affected by test set variance. We then ablated GrinAdapt with two modified baselines, CBMT \cite{tang2023source} with no integrated label and DPL \cite{chen2021source} with integrated label, and found that both the integrated label and designated teach-student scheme contributed to the improvement. We finally demonstrated the similar overall 5-domain artery and vein segmentation performance of GrInAdapt based on one single test image for each domain compared to the integrated label which requires 5 images from all domains in the test set, validating its consistent improvement on all domains and strong generalizability.
We further evaluated GrInAdapt across three sites (Table \ref{tab:both_only_site_assd_all_1deci}) on both Dice and ASSD metrics and again witnessed consistent improvements of Dice and reduced ASSD (on average 0.42) on all sites versus the source model. Compared to the integrated label, GrInAdapt also achieved comparable results on Artery and Vein segmentation and improved FAZ performance, demonstrating its robust and generalizable performance across sites.

%% file: dice_assd_table_results/dice_no_site_avf.tex
\begin{table}[t]
\centering
\setlength{\tabcolsep}{0pt}
\caption{Methods comparison on DSC score across various domains. A, V, F stand for artery, vein and FAZ. \textcolor{green!70!black}{$\Delta$} indicates improvements compared to the source model. Optic disc F is counted as \textcolor{red!70!black}{0 (fail)} / \textcolor{green!70!black}{100 (success)} if the output has FAZ / no FAZ prediction.}
\label{tab:dce_nosite_avf}
\small
{\fontsize{8pt}{10pt}\selectfont 
\begin{tabular}{c|c|ccc|c|c|c}
\hline
\multirow{4}{*}{Methods} & \multirow{4}{*}{C} & \multicolumn{6}{c}{\cite{tang2023source} DSC (\%) (mean $\pm$ standard deviation)  $\uparrow$} \\ \cline{3-8}
 & & \multicolumn{3}{c|}{Macula} & \multicolumn{1}{c|}{Optic Disc} & \multicolumn{1}{c|}{Macula} & \multirow{3}{*}{All} \\ \cline{3-7}
 & & Cirrus(D1) & Maestro2(D2) & Triton(D3) & Cirrus(D4) & Triton(D5) & \\ \cline{3-7}
 & & \multicolumn{4}{c|}{6$\times$6} & 12$\times$12 &  \\ 
\hline

\rowcolor{mediumGray}
Integrated & A
  & $74.9$$\pm$$6.9$ & $69.4$$\pm$$4.8$ & $71.2$$\pm$$7.9$ & $63.4$$\pm$$22.0$ & $65.5$$\pm$$6.6$ & $70.0$$\pm$$12.0$ \\
\cline{2-8}
\rowcolor{lightGray}
label & V
  & $77.9$$\pm$$9.9$ & $71.0$$\pm$$5.4$ & $75.1$$\pm$$6.4$ & $64.1$$\pm$$20.3$ & $70.9$$\pm$$4.2$ & $72.6$$\pm$$11.9$ \\
\cline{2-8}
\rowcolor{white}
(Fig. 1d) & F
  & $82.3$$\pm$$22.6$ & $83.5$$\pm$$22.0$ & $80.3$$\pm$$24.8$
  & \textcolor{green!70!black}{\textbf{$100$$\pm$$0.0$}}
  & $75.4$$\pm$$25.4$ & $81.6$$\pm$$23.2$ \\
\hline
\hline

\rowcolor{mediumGray}
Source & A
  & $70.0$$\pm$$6.4$ & $64.5$$\pm$$5.0$ & $65.0$$\pm$$9.0$ & $70.3$$\pm$$7.3$ & $58.5$$\pm$$5.9$ & $66.9$$\pm$$7.8$ \\
\cline{2-8}
\rowcolor{lightGray}
model & V
  & $71.3$$\pm$$9.0$ & $64.5$$\pm$$6.0$ & $68.3$$\pm$$8.2$ & $69.6$$\pm$$6.2$ & $62.6$$\pm$$5.0$ & $68.3$$\pm$$7.8$ \\
\cline{2-8}
\rowcolor{white}
($f_{\theta_S}$) & F
  & $79.4$$\pm$$30.2$ & $89.7$$\pm$$6.3$ & $83.1$$\pm$$23.8$
  & \textcolor{red!70!black}{$0$$\pm$$0.0$}
  & $76.5$$\pm$$30.4$ & $83.9$$\pm$$22.6$ \\
\hline

\rowcolor{mediumGray}
Ensemble & A
  & $71.7$$\pm$$6.2$ & $66.4$$\pm$$4.6$ & $67.0$$\pm$$9.3$ & $63.8$$\pm$$23.4$ & $60.6$$\pm$$7.3$ & $67.2$$\pm$$12.5$ \\
\cline{2-8}
\rowcolor{lightGray}
prediction & V
  & $73.5$$\pm$$9.3$ & $67.3$$\pm$$5.2$ & $71.1$$\pm$$7.8$ & $63.2$$\pm$$21.0$ & $65.9$$\pm$$5.2$ & $69.2$$\pm$$11.9$ \\
\cline{2-8}
\rowcolor{white}
(3 models) & F
  & $87.0$$\pm$$24.5$
  & \textbf{$90.7$$\pm$$6.2$}
  & $86.3$$\pm$$20.9$
  & \textcolor{red!70!black}{$0$$\pm$$0.0$}
  & \textbf{$83.8$$\pm$$23.1$} & $87.6$$\pm$$19.1$ \\
\hline

\rowcolor{mediumGray}
CBMT\cite{tang2023source} & A
  & $70.7$$\pm$$4.4$ & $62.8$$\pm$$5.7$ & $65.1$$\pm$$4.6$ & $71.1$$\pm$$4.4$ & $62.6$$\pm$$3.6$ & $66.9$$\pm$$5.7$ \\
\cline{2-8}
\rowcolor{lightGray}
w/ ensemble & V
  & $71.1$$\pm$$6.4$ & $63.7$$\pm$$5.0$ & $68.0$$\pm$$4.2$ & $69.7$$\pm$$5.0$ & $66.6$$\pm$$3.5$ & $68.0$$\pm$$5.6$ \\
\cline{2-8}
\rowcolor{white}
prediction & F
  & $0$$\pm$$0.0$ & $83.8$$\pm$$22.2$ & $42.7$$\pm$$45.4$
  & \textcolor{red!70!black}{$0$$\pm$$0.0$}
  & $4.2$$\pm$$16.6$ & $42.9$$\pm$$45.0$ \\
\hline

\rowcolor{mediumGray}
DPL\cite{chen2021source} w/ & A
  & \textbf{$73.3$$\pm$$4.9$} & $66.4$$\pm$$5.9$ & $67.8$$\pm$$8.8$
  & \textbf{$75.6$$\pm$$5.1$} & $61.7$$\pm$$6.9$ & $70.1$$\pm$$7.7$ \\
\cline{2-8}
\rowcolor{lightGray}
integrated & V
  & \textbf{$74.8$$\pm$$7.1$} & $66.3$$\pm$$6.7$ & $70.1$$\pm$$8.6$
  & \textbf{$73.7$$\pm$$5.0$} & $64.9$$\pm$$6.0$ & $70.9$$\pm$$7.8$ \\
\cline{2-8}
\rowcolor{white}
label & F
  & $88.1$$\pm$$8.7$ & $86.4$$\pm$$9.6$ & $84.2$$\pm$$18.4$
  & \textcolor{green!70!black}{\textbf{$100$$\pm$$0.0$}}
  & $78.7$$\pm$$22.4$ & $85.7$$\pm$$14.3$ \\
\hline
\hline
\rowcolor{mediumGray}
 & A
   & $73.1$$\pm$$5.3$\textcolor{green!70!black}{\scriptsize$\smallscript\Delta$3.1}
   & \textbf{$67.2$$\pm$$5.3$}\textcolor{green!70!black}{\scriptsize$\smallscript\Delta$2.7}
   & \textbf{$69.0$$\pm$$8.0$}\textcolor{green!70!black}{\scriptsize$\smallscript\Delta$4.0}
   & $74.4$$\pm$$5.9$\textcolor{green!70!black}{\scriptsize$\smallscript\Delta$4.1}
   & \textbf{$63.3$$\pm$$6.5$}\textcolor{green!70!black}{\scriptsize$\smallscript\Delta$4.8}
   & \textbf{$70.5$$\pm$$7.0$}\textcolor{green!70!black}{\scriptsize$\smallscript\Delta$3.6} \\
\cline{2-8}
\rowcolor{lightGray}
\textbf{GrInAdapt} & V
  & $74.6$$\pm$$7.5$\textcolor{green!70!black}{\scriptsize$\smallscript\Delta$3.3}
  & \textbf{$67.7$$\pm$$5.4$}\textcolor{green!70!black}{\scriptsize$\smallscript\Delta$3.2}
  & \textbf{$71.6$$\pm$$7.1$}\textcolor{green!70!black}{\scriptsize$\smallscript\Delta$3.3}
  & $73.5$$\pm$$5.5$\textcolor{green!70!black}{\scriptsize$\smallscript\Delta$3.9}
  & \textbf{$66.9$$\pm$$5.1$}\textcolor{green!70!black}{\scriptsize$\smallscript\Delta$4.3}
  & \textbf{$71.7$$\pm$$6.8$}\textcolor{green!70!black}{\scriptsize$\smallscript\Delta$3.4} \\
\cline{2-8}
\rowcolor{white}
\textbf{(Ours)} & F
  & \textbf{$90.9$$\pm$$6.2$}\textcolor{green!70!black}{\scriptsize$\smallscript\Delta$11.5}
  & $90.3$$\pm$$7.9$\textcolor{green!70!black}{\scriptsize$\smallscript\Delta$0.6}
  & \textbf{$87.3$$\pm$$17.6$}\textcolor{green!70!black}{\scriptsize$\smallscript\Delta$4.2}
  & \textcolor{green!70!black}{\textbf{$100$$\pm$$0.0$}}
  & $83.5$$\pm$$23.1$\textcolor{green!70!black}{\scriptsize$\smallscript\Delta$7.0}
  & \textbf{$89.0$$\pm$$13.2$}\textcolor{green!70!black}{\scriptsize$\smallscript\Delta$5.1} \\
\hline

\end{tabular}
}
\end{table}

%% file: dice_assd_table_results/both_only_site_assd_all_1deci.tex
\begin{table}[t]
\setlength{\tabcolsep}{0.5pt}
\centering
\caption{Cross-site methods comparison on the DSC score and ASSD distances.}
\label{tab:both_only_site_assd_all_1deci}
\small
{\fontsize{8pt}{10pt}\selectfont
\begin{tabular}{c|c|ccc|ccc|c} 
\hline
\multirow{2}{*}{Methods} & \multirow{2}{*}{C}
  & \multicolumn{3}{c|}{\cite{tang2023source} DSC (\%) $\uparrow$}
  & \multicolumn{4}{c}{\cite{tang2023source} ASSD (pixel) $\downarrow$} \\ 
\cline{3-9}
& & UW & UAB & UCSD & UW & UAB & UCSD & All \\
\hline

\rowcolor{mediumGray}
Inegrated & A 
    & $70.2$$\pm$$7.8$ & $74.3$$\pm$$6.7$ & $65.7$$\pm$$17.8$
    & $1.4$$\pm$$0.7$ & $1.1$$\pm$$0.5$ & $1.9$$\pm$$1.8$ & $1.5$$\pm$$1.2$ \\
\cline{2-9}
\rowcolor{lightGray}
Label & V
    & $70.7$$\pm$$7.6$ & $78.1$$\pm$$6.1$ & $69.6$$\pm$$17.8$
    & $1.3$$\pm$$0.6$ & $1.0$$\pm$$0.5$ & $1.7$$\pm$$1.7$ & $1.3$$\pm$$1.1$ \\
\cline{2-9}
\rowcolor{white}
(Fig. 1d)  & F
    & $73.9$$\pm$$33.8$ & $84.8$$\pm$$14.0$ & $88.4$$\pm$$5.4$
    & $1.2$$\pm$$0.7$ & $1.7$$\pm$$2.5$ & $1.3$$\pm$$0.6$ & $1.4$$\pm$$1.5$ \\
\hline
\hline

\rowcolor{mediumGray}
Source & A
    & $67.0$$\pm$$7.8$ & $70.0$$\pm$$7.2$ & $63.8$$\pm$$7.5$
    & $2.0$$\pm$$1.2$ & $1.5$$\pm$$0.7$ & $2.2$$\pm$$1.3$ & $1.9$$\pm$$1.1$ \\
\cline{2-9}
\rowcolor{lightGray}
Model & V
    & $66.7$$\pm$$7.2$ & $72.3$$\pm$$6.7$ & $66.6$$\pm$$8.5$
    & $1.9$$\pm$$1.1$ & $1.6$$\pm$$1.0$ & $2.3$$\pm$$1.8$ & $1.9$$\pm$$1.3$ \\
\cline{2-9}
\rowcolor{white}
($f_{\theta_S}$) & F
    & $74.1$$\pm$$33.4$ & $90.6$$\pm$$4.9$ & $89.9$$\pm$$7.2$
    & $1.5$$\pm$$2.4$ & $0.9$$\pm$$0.7$ & $1.0$$\pm$$0.7$ & $1.2$$\pm$$1.6$ \\
\hline

\rowcolor{mediumGray}
\textbf{GrIn}& A
    & \textbf{$70.3$$\pm $$7.3$}\textcolor{green!70!black}{\scriptsize$\smallscript\Delta$3.3}
      & \textbf{$73.3$$\pm$$6.7$}\textcolor{green!70!black}{\scriptsize$\smallscript\Delta$3.3}
      & \textbf{$68.1$$\pm$$6.2$}\textcolor{green!70!black}{\scriptsize$\smallscript\Delta$4.3}
    & \textbf{$1.6$$\pm$$0.8$}\textcolor{green!70!black}{\scriptsize$\smallscript\Delta$.4}
      & \textbf{$1.2$$\pm$$0.5$}\textcolor{green!70!black}{\scriptsize$\smallscript\Delta$.3}
      & \textbf{$1.7$$\pm$$0.9$}\textcolor{green!70!black}{\scriptsize$\smallscript\Delta$.5}
      & \textbf{$1.5$$\pm$$0.8$}\textcolor{green!70!black}{\scriptsize$\smallscript\Delta$.4} \\
\cline{2-9}
\rowcolor{lightGray}
\textbf{Adapt} & V
    & \textbf{$70.3$$\pm$$6.6$}\textcolor{green!70!black}{\scriptsize$\smallscript\Delta$3.6}
      & \textbf{$75.1$$\pm$$5.9$}\textcolor{green!70!black}{\scriptsize$\smallscript\Delta$2.8}
      & \textbf{$70.2$$\pm$$7.1$}\textcolor{green!70!black}{\scriptsize$\smallscript\Delta$3.6}
    & \textbf{$1.4$$\pm$$0.7$}\textcolor{green!70!black}{\scriptsize$\smallscript\Delta$.5}
      & \textbf{$1.1$$\pm$$0.6$}\textcolor{green!70!black}{\scriptsize$\smallscript\Delta$.5}
      & \textbf{$1.6$$\pm$$0.9$}\textcolor{green!70!black}{\scriptsize$\smallscript\Delta$.7}
      & \textbf{$1.4$$\pm$$0.8$}\textcolor{green!70!black}{\scriptsize$\smallscript\Delta$.5} \\
\cline{2-9}
\rowcolor{white}
\textbf{(Ours)} & F
    & \textbf{$84.7$$\pm$$20.$}\textcolor{green!70!black}{\scriptsize$\smallscript\Delta$10.6}
      & \textbf{$92.2$$\pm$$3.5$}\textcolor{green!70!black}{\scriptsize$\smallscript\Delta$1.6}
      & \textbf{$91.5$$\pm$$5.0$}\textcolor{green!70!black}{\scriptsize$\smallscript\Delta$1.6}
    & \textbf{$0.9$$\pm$$0.4$}\textcolor{green!70!black}{\scriptsize$\smallscript\Delta$.6}
      & \textbf{$0.7$$\pm$$0.3$}\textcolor{green!70!black}{\scriptsize$\smallscript\Delta$.2}
      & \textbf{$0.9$$\pm$$0.5$}\textcolor{green!70!black}{\scriptsize$\smallscript\Delta$.1}
      & \textbf{$0.8$$\pm$$0.4$}\textcolor{green!70!black}{\scriptsize$\smallscript\Delta$.4} \\
\hline

\end{tabular}
}
\end{table}

%% file: conclusion.tex
We have presented GrInAdapt, a novel framework for multi-target domain adaptation in retinal vessel segmentation. Using multiview and multimodal imaging, GrInAdapt refines pseudo-labels via three steps: grounding via registration, integrating multiple predictions with region-specific fusion, and adapting the source model using a teacher–student framework. Experiments on the large-scale AI-READI dataset demonstrate the superior label consensus and improved segmentation performance across diverse imaging domains achieved by GrInAdapt.
In future work, we plan to explore advanced registration techniques and incorporate additional modalities, with the goal of reducing annotation burdens and developing a fully automated, clinically deployable retinal vessel analysis system.

%% file: Supp/Supp_result.tex
\newcommand{\centeredappendix}{
    \clearpage
    \begin{center}
        {\Large\textbf{Supplementary Materials}} 
    \end{center}
}
\appendix
\centeredappendix

In this supplementary materials, we introduce more experimental results, illustration, and implementation details of GrInAdapt.

\begin{figure}
    \centering
    \setlength{\abovecaptionskip}{-3pt}  
    \setlength{\belowcaptionskip}{-3pt}    
    \includegraphics[width=\linewidth]{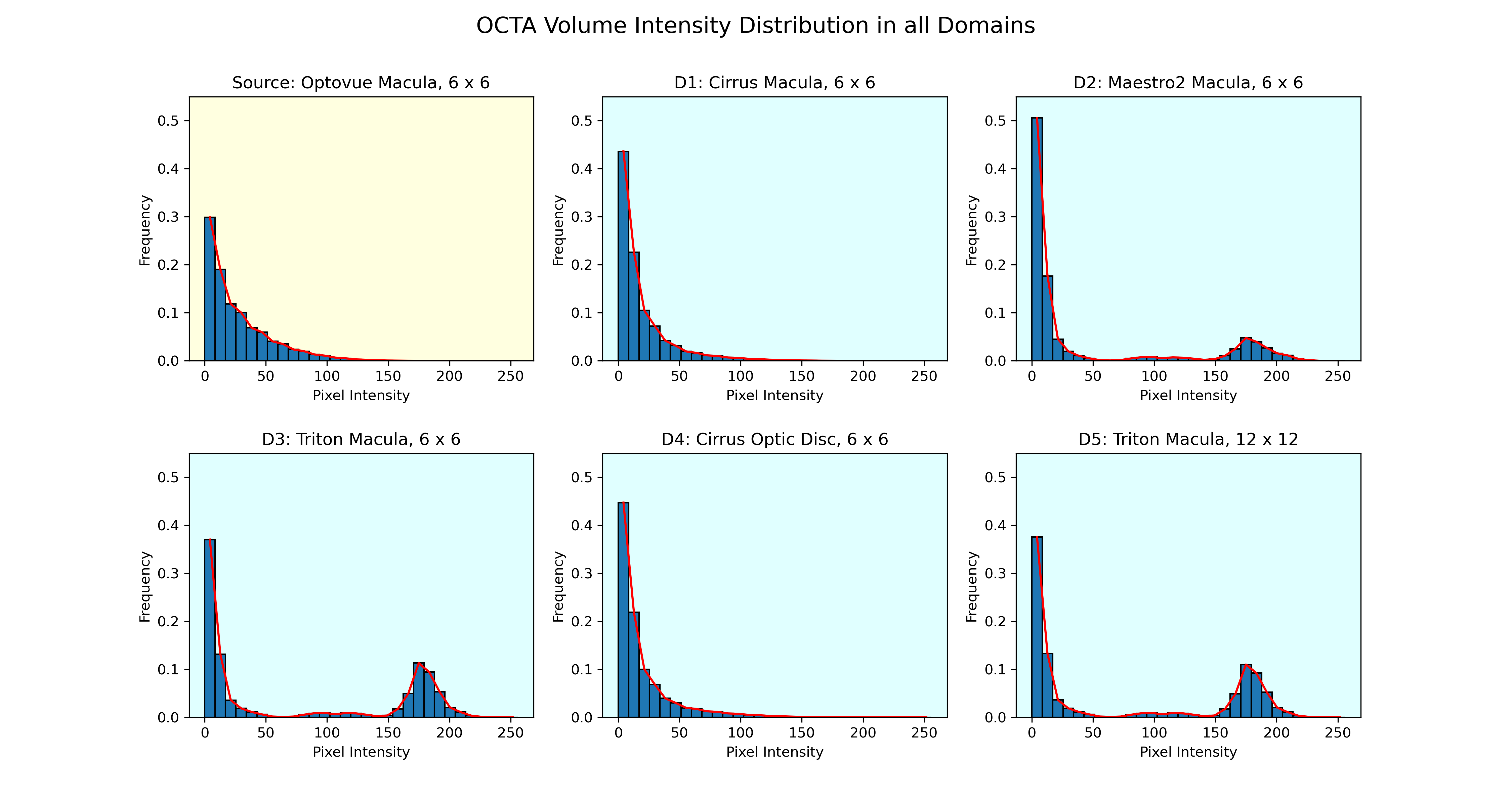}
    \caption{The intensity distribution for OCTA volumes from different domains (\textbf{a}, source domain, \textbf{b-f}, target domains). The frequency is accumulated and normalized across the test set from the test set of OCTA-500 (50 samples) and AI-READI datasets (80 samples) used for testing the source model and evaluating the GrInAdapt performace. Images pictured by Zeiss Cirrus machine have more simialr intensity distribution to the source Optovue domain. Images from Topcon machine (Maestro2 and Triton) have more different distribution. This indicates the performance gap between Zeiss Cirrus domains and Topcon machine domains when evaluating the source model or the adapted model.}
    \label{fig:input_dist}
\end{figure}

\section{Illustration of performance variance across domains} We included Figure \ref{fig:input_dist} to demonstrate the normalized intensity distributions of OCTA volumes from both the source and target domains. Clearly, both D1 and D4 imaged with Zeiss Cirrus device had a much more similar input intensity distribution to the source domain Optovue RTVue than the other domains This explains the higher performance on D1 and D4 compared to the other domains (D2, D3, D5).
\input{Supp/stat_table/scan_data}

\section{More Details of the AI-READI datset}


Table \ref{tab:scan_data} demonstrates more metadata of the images in the AI-READI
dataset. The 3D volumnes and 2D projection maps differ in shape for most of the domains. Table \ref{tab:disease_dist} shows the disease distribution of Type-2 Diabete Mellitus (T2Dm) in the AI-READI dataset. Patient were split into four rough groups: healthy, insulin dependent, pre-diabetic lifestyle, and taking oral or non-insulin injectable medicine. Of the 1060 patients with both OCT and OCTA volumes, 369 are healthy, 129 are insulin dependent, 242 live a pre-diabetic lifestyle, and 320 take oral or non-insulin injectable medication.

\section{Additional Quantitative Results}

In this section, we present more details of the experimental results. Although the main paper reported key performance metrics, here we provide a more-detailed breakdown across different frameworks, domains, and sites, as well as an evaluation on two additional baselines—GAN and NoNorm—on our test sets (More details of these two models are provided below). We compared GrInAdapt with two further baselines to fully assess its effectiveness on DSC (see Table \ref{tab:supp_dce_nosite_avf}) and ASSD (see Table \ref{tab:supp_assd_nosite_assd_1f}). Additionally, we compared GrInAdapt against these two baselines across three sites on both DSC and ASSD (see Table \ref{tab:supp_both_only_site_assd_all_1deci}). These results further reinforce our claims regarding the robustness and generalizability of GrInAdapt under diverse retinal imaging conditions.

\input{Supp/stat_table/disease_dist}

\section{More visualizations}
\begin{figure}
    \centering
    \includegraphics[width=1\linewidth]{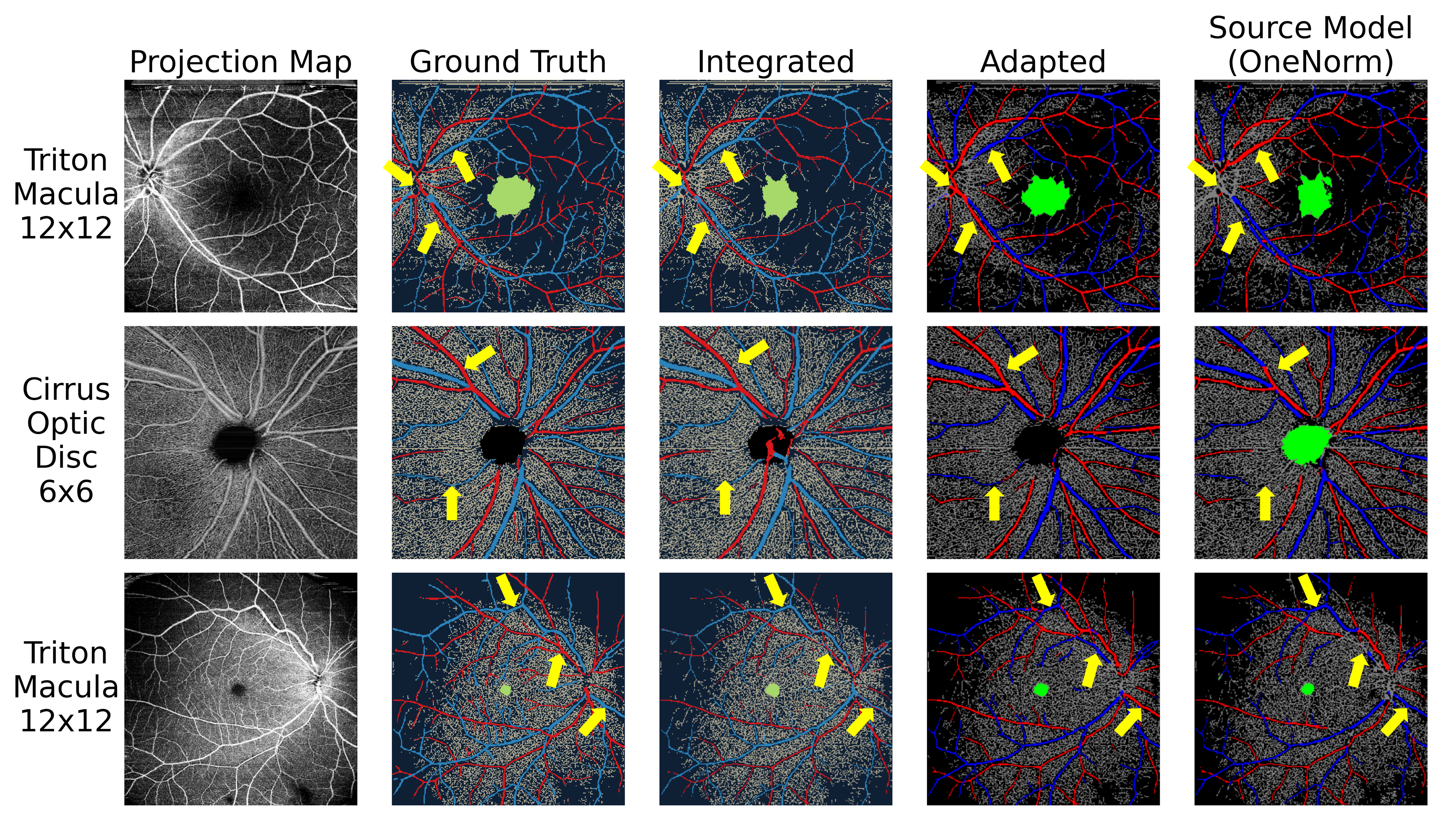}
    \caption{2D \textit{en face} projection map, ground truth label, integrated label, adapted model prediction, and source model prediction for four different eyes. The first two rows highlight eyes where the Adapted Model successfully improved on the source model, and the bottom row highlights an eye where the Adapted model did not fully improve / made more mistakes.}
    \label{fig:quali}
\end{figure}

In this section, we provide more qualitative examples to show the effectiveness of our GrInAdapt framework, as shown in Figure \ref{fig:quali}. 2 success and 1 failure cases of our GrInAdapted model are shown for a fair understanding. In the first two rows, it is clear that the adapted model is able to correct the source model's mistakes and get a result much closer to the integrated and ground-truth prediction for all the areas to which the yellow arrow points. In the bottom row, there are issues with the adapted model for areas to which the yellow arrows point, indicating places for further improvement.

%% file: Supp/stat_table/scan_data.tex
\begin{table}[ht]
    \centering
    \begin{tabular}{l|l|l|l|l|c|c}
        \hline
    \multirow{2}{*}{\textbf{Domain}} & \textbf{Manu-} & \multirow{2}{*}{\textbf{Device}} & \multirow{2}{*}{\textbf{Region}} & \multirow{2}{*}{\textbf{FOV}} & \multicolumn{2}{c}{\textbf{Shape}} \\ 
    \cline{6-7}
    & \textbf{facturer}& & & & \textbf{3D Volume} & \textbf{2D Proj. Map} \\
        \hline
        D1 & Zeiss & Cirrus & Macula & 6 x 6   & 1024 × 350 × 350 & 1024 × 1024 \\
        \hline
        D2 & Topcon & Maestro2 & Macula & 6 x 6  & 885 × 360 × 360  & 720 × 720 \\
        \hline
        D3 & Topcon & Triton & Macula & 6 x 6   & 992 × 320 × 320  & 640 × 640 \\
        \hline
        D4 & Zeiss & Cirrus & Optic Disc & 6 x 6   & 1024 × 350 × 350 & 1024 × 1024 \\
        \hline
        D5 & Topcon & Triton & Macula &  12 x 12 & 992 × 512 × 512  & 1024 × 1024 \\
        \hline

    \end{tabular}
    \vspace{2px}
    \caption{Metadata of the OCTA and OCT volumes and their corresponding 2D projection maps across different domains. FOV stands for field of view. Region denotes the center of the images.}
    \label{tab:scan_data}
\end{table}

%% file: Supp/stat_table/disease_dist.tex
\begin{table}[ht]
    \centering
    {\fontsize{9pt}{11pt}\selectfont
        \begin{tabular}{c|c||c|c|c|c}
            \hline
    \multirow{3}{*}{\textbf{Site}} & \multicolumn{4}{c|}{\textbf{Diabetes Status (T2DM) [count/\%]}} & \multirow{3}{*}{\textbf{Total}} \\ 
    \cline{2-5}
    &  \multirow{2}{*}{\textbf{Healthy}} & \textbf{Insulin} & \textbf{Pre-Diabetic} & \textbf{Oral/Non-} & \\  
    &  & \textbf{Dependent} & \textbf{Lifestyle} & \textbf{Insulin-Injectable} & \\
    \hline

            \hline
            \textbf{UW} & 160 & 34 & 97 & 92 & 383
            \\
            \hline
            \textbf{USCD} & 83 & 33 & 74 & 99 & 289
            \\
            \hline
            \textbf{UAB} & 126 & 62 & 71 & 129 & 388
            \\
            \hline
            \hline
            \textbf{Total} & 369 & 129 & 242 & 320 & 1060
        \end{tabular}
    }
    \vspace{2px}
    \caption{The disease distribution of Type-2 Diabete Mellitus (T2DM) in the AI-READI dataset. }
    \label{tab:disease_dist}
\end{table}

%% file: Supp/Supp_mat.tex
\section{Model Architecture of $f_{\theta_S}$}

\input{dice_assd_table_results/supp_dice_no_site}
\input{dice_assd_table_results/supp_assd_no_site_assd_1f}
\input{dice_assd_table_results/supp_both_site_assd_1f}

We utilized three models from two distinct architectures: a IPN-v2 \cite{li2024octa} like model
$f_{{\theta_S}}^{OneNorm}$, and a bigger model $f_{\theta_{S}}^{GAN}$ with GAN module borrowed from \cite{zhou2021learning}.  

\noindent\textbf{Architecture of the \textit{OneNorm }model.} The $f_{\theta_{S}}^{OneNorm}$ architecture builds upon the framework proposed in IPNv2 \cite{li2024octa}, with several key modifications to enhance feature extraction and generalization. 
First, we incorporated a 2D modified U-Net \cite{ronneberger2015u} branch to extract features from the 2D projection maps. Second, we enlarged the consistent number of input channels $C=128$ for each convolution block used in \cite{li2024octa}.
Specifically, we used:
$$C \rightarrow 2C \rightarrow 2C \rightarrow 4C \rightarrow 4C$$ 
Third, we added an extra GroupNorm layer at the end of each double convolutional block. We also tested a \textit{NoNorm} variant and demonstrated its result in \ref{tab:supp_dce_nosite_avf}.
Lastly, we replaced the original double convolutional out-layer with a single $1 \times 1$ convolution.

After feature extraction, the 2D features from the projection maps are concatenated with the features extracted from the corresponding 3D OCT and OCTA volumes. The combined feature set is then passed through a U-Net head for final segmentation.

\noindent\textbf{Architecture of the \textit{GAN} model.} This model was build upon the \textit{OneNorm} model, but was also equipped with the GAN module proposed in \cite{zhou2021learning}. We added two GroupNorm layers within each double convolutional block (one after each convolution), and replaced the BatchNorm layers in the side block (as described in \cite{zhou2021learning}) with GroupNorm to better accommodate the variability in data/pixel-value distributions across retinal imaging devices. We also posted the results of this model in Table \ref{tab:supp_dce_nosite_avf}.

\section{More Details of Source Model Training}

\noindent\textbf{Input and output format of OCTA CAVF segmentation.} We first introduce the input and output format.
Given a data sample $x_{\text{octa}}\in\mathcal{X}_{\text{OCTA}}$ serving as the input of $f_{\theta_S}$, let $(H_{3d}, W_{3d}, D_{3d})$ and $(H_{2d}, W_{2d})$ be the height, width and depth (only for 3D volumes) of 3D volumes and 2D projection maps. $x_{\text{octa}}$ consists of two 3D volumes (denoted with subscript $V$) and two 2D \textit{en face} projection maps (denoted with subscript $P$):  
$$x_{octa} = \{x_{\text{flow}_V}, x_{\text{structural}_V}, x_{\text{flow}_P}, x_{\text{structural}_P}\}$$
where $x_{\text{flow}_V} \in \mathbf{R}^{H_{3d}\times W_{3d}\times D_{3d}}$ and $ x_{\text{flow}_P} \in \mathbf{R}^{H_{2d}\times W_{2d}}$ denote the OCTA flow volume and its superficial projection maps, $x_{\text{structural}_V} \in \mathbf{R}^{H_{3d}\times W_{3d}\times D_{3d}}$ and $ x_{\text{structural}_P} \in \mathbf{R}^{H_{2d}\times W_{2d}}$ denote the structural OCT volumes and its superficial projection maps.
The output $y_{cavf} \in \mathbf{R}^{H\times W\times 5}$ is a 5-channel CAVF probability map, with 0: Background (BG), 1: Capillary (C), 2: Artery (A), 3: Vein (V), 4: Foveal Avascular Zone (FAZ or F). We directly input the whole images into the model instead of input image patches to better ensure the location information were correctly processed. Despite raw data from different domains may have domain-specific H, W and D, we set $H=W=256$ and $D=128$ and resized all the data to this shape for a fair comparison. 

\noindent\textbf{Statistics of the OCTA-500 dataset.} Among the 300 samples from the OCTA-500 dataset used to train the source model, 69.7\% had ophthalmic diseases --- including AMD, DR, CNV, CSC, RVO, and others (where ‘others’ here refers to diseases with less than 8 samples e.g. retinal detachment, retinal hemorrhage, optic atrophy, epiretinal membrane, retinitis pigmentosa, central retinal artery occlusion, retinoschisis, etc). The shape of the raw 3D OCTA and OCT volume is (640, 400, 400), and (400, 400) for the corresponding 2D preojction maps.

\noindent\textbf{Data augmentation.} For each subject with a 3D OCTA / 3D OCT / 2D OCTA / 2D OCT images, random horizontal flips, vertical flips, rotations by multiples of 90°, and z-axis shifts maybe be applied, with consistent parameter shared across 4 modalities to make sure the augmentation is realistic and consistent on all images in the bag. The probability is set to be 0.25 and independently sampled for each of the operations.

\noindent\textbf{Training.}The models were trained for 300 epochs. The batch size is set to be 4 for OneNorm and NoNorm, but 1 for the GAN model due to its larger architecture. The OneNorm model employed an Adam optimizer with a 0.0001 learning rate and beta values of 0.9 and 0.99. The GAN model used multiple Adam optimizers (following the training code from \cite{zhou2021learning}) with the same settings as the OneNorm model.
Training is performed using the PyTorch framework on an NVIDIA RTX A6000 GPU. 

\noindent\textbf{Ensemble prediction.}
For each data sample in the target domains, we generated three CAVF probability maps using the \textit{OneNorm}, \textit{NoNorm} and \textit{GAN} model. These probability maps were then averaged to get the ensemble prediction. We also used this ensemble prediction to serve as the CAVF map for integration.

%% file: dice_assd_table_results/supp_dice_no_site.tex
\begin{table}[t]
\centering
\setlength{\tabcolsep}{0pt}
\caption{Methods comparison on DSC score across various domains. A, V, F stand for artery, vein and FAZ. \textcolor{green!70!black}{$\Delta$} indicates improvements compared to the source model. Optic disc F is counted as \textcolor{red!70!black}{0 (fail)} / \textcolor{green!70!black}{100 (success)} if the output has FAZ / no FAZ prediction.}
\label{tab:supp_dce_nosite_avf}
\small
{\fontsize{8pt}{10pt}\selectfont 
\begin{tabular}{c|c|ccc|c|c|c}
\hline
\multirow{4}{*}{Methods} & \multirow{4}{*}{C} & \multicolumn{6}{c}{\cite{tang2023source} DSC (\%) (mean $\pm$ standard deviation)  $\uparrow$} \\ \cline{3-8}
 & & \multicolumn{3}{c|}{Macula} & \multicolumn{1}{c|}{Optic Disc} & \multicolumn{1}{c|}{Macula} & \multirow{3}{*}{All} \\ \cline{3-7}
 & & Cirrus(D1) & Maestro2(D2) & Triton(D3) & Cirrus(D4) & Triton(D5) & \\ \cline{3-7}
 & & \multicolumn{4}{c|}{6$\times$6} & 12$\times$12 &  \\ 
\hline

\rowcolor{mediumGray}
Integrated & A
  & $74.9$$\pm$$6.9$ & $69.4$$\pm$$4.8$ & $71.2$$\pm$$7.9$ & $63.4$$\pm$$22.0$ & $65.5$$\pm$$6.6$ & $70.0$$\pm$$12.0$ \\
\cline{2-8}
\rowcolor{lightGray}
label & V
  & $77.9$$\pm$$9.9$ & $71.0$$\pm$$5.4$ & $75.1$$\pm$$6.4$ & $64.1$$\pm$$20.3$ & $70.9$$\pm$$4.2$ & $72.6$$\pm$$11.9$ \\
\cline{2-8}
\rowcolor{white}
(Fig. 1d) & F
  & $82.3$$\pm$$22.6$ & $83.5$$\pm$$22.0$ & $80.3$$\pm$$24.8$
  & \textcolor{green!70!black}{\textbf{$100$$\pm$$0.0$}}
  & $75.4$$\pm$$25.4$ & $81.6$$\pm$$23.2$ \\
\hline
\hline

\rowcolor{mediumGray}
Source & A
  & $70.0$$\pm$$6.4$ & $64.5$$\pm$$5.0$ & $65.0$$\pm$$9.0$ & $70.3$$\pm$$7.3$ & $58.5$$\pm$$5.9$ & $66.9$$\pm$$7.8$ \\
\cline{2-8}
\rowcolor{lightGray}
model & V
  & $71.3$$\pm$$9.0$ & $64.5$$\pm$$6.0$ & $68.3$$\pm$$8.2$ & $69.6$$\pm$$6.2$ & $62.6$$\pm$$5.0$ & $68.3$$\pm$$7.8$ \\
\cline{2-8}
\rowcolor{white}
($f_{\theta_S}$) & F
  & $79.4$$\pm$$30.2$ & $89.7$$\pm$$6.3$ & $83.1$$\pm$$23.8$
  & \textcolor{red!70!black}{$0$$\pm$$0.0$}
  & $76.5$$\pm$$30.4$ & $83.9$$\pm$$22.6$ \\
\hline

\rowcolor{mediumGray}
NoNorm & A
  & $67.5$$\pm$$7.2$ & $64.7$$\pm$$4.8$ & $65.1$$\pm$$9.7$ & $67.3$$\pm$$8.2$ & $58.1$$\pm$$7.1$ & $65.9$$\pm$$8.0$ \\
\cline{2-8}
\rowcolor{lightGray}
model & V
  & $68.6$$\pm$$10.8$ & $66.2$$\pm$$5.3$ & $69.9$$\pm$$7.6$ & $65.9$$\pm$$7.5$ & $64.6$$\pm$$4.6$ & $68.1$$\pm$$8.0$ \\
\cline{2-8}
\rowcolor{white}
 & F
  & $89.4$$\pm$$11.2$ & $88.2$$\pm$$6.9$ & $82.5$$\pm$$23.7$
  & \textcolor{red!70!black}{$0$$\pm$$0.0$}
  & $81.2$$\pm$$22.6$ & $85.8$$\pm$$17.9$ \\
\hline

\rowcolor{mediumGray}
GAN & A
  & $70.9$$\pm$$6.1$ & $64.4$$\pm$$6.6$ & $64.9$$\pm$$9.7$ & $62.9$$\pm$$22.1$ & $58.0$$\pm$$6.8$ & $65.6$$\pm$$12.3$ \\
\cline{2-8}
\rowcolor{lightGray}
model & V
  & $72.4$$\pm$$7.8$ & $66.0$$\pm$$6.1$ & $68.4$$\pm$$8.8$ & $62.1$$\pm$$20.3$ & $62.0$$\pm$$6.0$ & $67.4$$\pm$$11.8$ \\
\cline{2-8}
\rowcolor{white}
& F
  & $83.4$$\pm$$23.7$ & $85.5$$\pm$$22.7$ & $82.8$$\pm$$20.6$
  & \textcolor{red!70!black}{$0$$\pm$$0.0$}
  & $75.5$$\pm$$25.4$ & $83.7$$\pm$$21.6$ \\
\hline

\rowcolor{mediumGray}
Ensemble & A
  & $71.7$$\pm$$6.2$ & $66.4$$\pm$$4.6$ & $67.0$$\pm$$9.3$ & $63.8$$\pm$$23.4$ & $60.6$$\pm$$7.3$ & $67.2$$\pm$$12.5$ \\
\cline{2-8}
\rowcolor{lightGray}
prediction & V
  & $73.5$$\pm$$9.3$ & $67.3$$\pm$$5.2$ & $71.1$$\pm$$7.8$ & $63.2$$\pm$$21.0$ & $65.9$$\pm$$5.2$ & $69.2$$\pm$$11.9$ \\
\cline{2-8}
\rowcolor{white}
(3 models) & F
  & $87.0$$\pm$$24.5$
  & \textbf{$90.7$$\pm$$6.2$}
  & $86.3$$\pm$$20.9$
  & \textcolor{red!70!black}{$0$$\pm$$0.0$}
  & \textbf{$83.8$$\pm$$23.1$} & $87.6$$\pm$$19.1$ \\
\hline

\rowcolor{mediumGray}
CBMT\cite{tang2023source} & A
  & $70.7$$\pm$$4.4$ & $62.8$$\pm$$5.7$ & $65.1$$\pm$$4.6$ & $71.1$$\pm$$4.4$ & $62.6$$\pm$$3.6$ & $66.9$$\pm$$5.7$ \\
\cline{2-8}
\rowcolor{lightGray}
w/ ensemble & V
  & $71.1$$\pm$$6.4$ & $63.7$$\pm$$5.0$ & $68.0$$\pm$$4.2$ & $69.7$$\pm$$5.0$ & $66.6$$\pm$$3.5$ & $68.0$$\pm$$5.6$ \\
\cline{2-8}
\rowcolor{white}
prediction & F
  & $0$$\pm$$0.0$ & $83.8$$\pm$$22.2$ & $42.7$$\pm$$45.4$
  & \textcolor{red!70!black}{$0$$\pm$$0.0$}
  & $4.2$$\pm$$16.6$ & $42.9$$\pm$$45.0$ \\
\hline

\rowcolor{mediumGray}
DPL\cite{chen2021source} w/ & A
  & \textbf{$73.3$$\pm$$4.9$} & $66.4$$\pm$$5.9$ & $67.8$$\pm$$8.8$
  & \textbf{$75.6$$\pm$$5.1$} & $61.7$$\pm$$6.9$ & $70.1$$\pm$$7.7$ \\
\cline{2-8}
\rowcolor{lightGray}
integrated & V
  & \textbf{$74.8$$\pm$$7.1$} & $66.3$$\pm$$6.7$ & $70.1$$\pm$$8.6$
  & \textbf{$73.7$$\pm$$5.0$} & $64.9$$\pm$$6.0$ & $70.9$$\pm$$7.8$ \\
\cline{2-8}
\rowcolor{white}
label & F
  & $88.1$$\pm$$8.7$ & $86.4$$\pm$$9.6$ & $84.2$$\pm$$18.4$
  & \textcolor{green!70!black}{\textbf{$100$$\pm$$0.0$}}
  & $78.7$$\pm$$22.4$ & $85.7$$\pm$$14.3$ \\
\hline
\hline
\rowcolor{mediumGray}
 & A
   & $73.1$$\pm$$5.3$\textcolor{green!70!black}{\scriptsize$\smallscript\Delta$3.1}
   & \textbf{$67.2$$\pm$$5.3$}\textcolor{green!70!black}{\scriptsize$\smallscript\Delta$2.7}
   & \textbf{$69.0$$\pm$$8.0$}\textcolor{green!70!black}{\scriptsize$\smallscript\Delta$4.0}
   & $74.4$$\pm$$5.9$\textcolor{green!70!black}{\scriptsize$\smallscript\Delta$4.1}
   & \textbf{$63.3$$\pm$$6.5$}\textcolor{green!70!black}{\scriptsize$\smallscript\Delta$4.8}
   & \textbf{$70.5$$\pm$$7.0$}\textcolor{green!70!black}{\scriptsize$\smallscript\Delta$3.6} \\
\cline{2-8}
\rowcolor{lightGray}
\textbf{GrInAdapt} & V
  & $74.6$$\pm$$7.5$\textcolor{green!70!black}{\scriptsize$\smallscript\Delta$3.3}
  & \textbf{$67.7$$\pm$$5.4$}\textcolor{green!70!black}{\scriptsize$\smallscript\Delta$3.2}
  & \textbf{$71.6$$\pm$$7.1$}\textcolor{green!70!black}{\scriptsize$\smallscript\Delta$3.3}
  & $73.5$$\pm$$5.5$\textcolor{green!70!black}{\scriptsize$\smallscript\Delta$3.9}
  & \textbf{$66.9$$\pm$$5.1$}\textcolor{green!70!black}{\scriptsize$\smallscript\Delta$4.3}
  & \textbf{$71.7$$\pm$$6.8$}\textcolor{green!70!black}{\scriptsize$\smallscript\Delta$3.4} \\
\cline{2-8}
\rowcolor{white}
\textbf{(Ours)} & F
  & \textbf{$90.9$$\pm$$6.2$}\textcolor{green!70!black}{\scriptsize$\smallscript\Delta$11.5}
  & $90.3$$\pm$$7.9$\textcolor{green!70!black}{\scriptsize$\smallscript\Delta$0.6}
  & \textbf{$87.3$$\pm$$17.6$}\textcolor{green!70!black}{\scriptsize$\smallscript\Delta$4.2}
  & \textcolor{green!70!black}{\textbf{$100$$\pm$$0.0$}}
  & $83.5$$\pm$$23.1$\textcolor{green!70!black}{\scriptsize$\smallscript\Delta$7.0}
  & \textbf{$89.0$$\pm$$13.2$}\textcolor{green!70!black}{\scriptsize$\smallscript\Delta$5.1} \\
\hline

\end{tabular}
}
\end{table}

%% file: dice_assd_table_results/supp_assd_no_site_assd_1f.tex
\begin{table}[t]
\centering
\setlength{\tabcolsep}{1pt}
\caption{Full results of methods comparison on ASSD score across various domains. A, V, F stand for artery, vein and FAZ. \textcolor{green!70!black}{$\Delta$} indicates improvements compared to the source model. Optic disc ASSD is all set to be N/A because no ground truth FAZ area will appear.}
\label{tab:supp_assd_nosite_assd_1f}
\small
{\fontsize{8pt}{10pt}\selectfont 
\begin{tabular}{c|c|ccc|c|c|c}
\hline
\multirow{4}{*}{Methods} & \multirow{4}{*}{C} & \multicolumn{6}{c}{\cite{tang2023source} ASSD (pixel) (mean $\pm$ standard deviation)  $\downarrow$} \\ \cline{3-8}
 & & \multicolumn{3}{c|}{Macula} & \multicolumn{1}{c|}{Optic Disc} & \multicolumn{1}{c|}{Macula} & \multirow{3}{*}{All} \\ \cline{3-7}
 & & Cirrus(D1) & Maestro2(D2) & Triton(D3) & Cirrus(D4) & Triton(D5) & \\ \cline{3-7}
 & & \multicolumn{4}{c|}{6$\times$6} & 12$\times$12 &  \\ 
\hline

\rowcolor{mediumGray}
Integrated & A
  & $1.2$$\pm$$0.6$ & $1.1$$\pm$$0.4$ & $1.4$$\pm$$0.7$ & $2.3$$\pm$$2.1$ & $1.9$$\pm$$0.6$ & $1.5$$\pm$$1.2$ \\
\cline{2-8}
\rowcolor{lightGray}
label & V
  & $1.0$$\pm$$0.6$ & $1.0$$\pm$$0.4$ & $1.2$$\pm$$0.6$ & $2.4$$\pm$$1.9$ & $1.5$$\pm$$0.4$ & $1.3$$\pm$$1.1$ \\
\cline{2-8}
\rowcolor{white}
(Fig. 1d) & F
  & $1.5$$\pm$$0.7$ & $1.4$$\pm$$0.7$ & $1.4$$\pm$$2.0$
  & N/A
  & $1.7$$\pm$$2.8$ & $1.4$$\pm$$1.5$ \\
\hline
\hline

\rowcolor{mediumGray}
Source & A
  & $1.7$$\pm$$1.2$ & $1.7$$\pm$$0.9$ & $2.1$$\pm$$1.2$ & $1.9$$\pm$$1.2$ & $2.9$$\pm$$1.0$ & $1.9$$\pm$$1.1$ \\
\cline{2-8}
\rowcolor{lightGray}
model & V
  & $1.5$$\pm$$1.1$ & $2.0$$\pm$$2.0$ & $2.0$$\pm$$1.2$ & $2.1$$\pm$$0.9$ & $2.8$$\pm$$0.9$ & $1.9$$\pm$$1.3$ \\
\cline{2-8}
\rowcolor{white}
($f_{\theta_S}$) & F
  & $1.8$$\pm$$2.9$ & $1.0$$\pm$$0.5$ & $0.9$$\pm$$0.7$
  & N/A
  & $0.9$$\pm$$0.7$ & $1.2$$\pm$$1.6$ \\
\hline

\rowcolor{mediumGray}
NoNorm & A
  & $2.3$$\pm$$1.9$ & $1.7$$\pm$$0.7$ & $2.2$$\pm$$1.3$ & $2.4$$\pm$$1.2$ & $3.0$$\pm$$1.1$ & $2.2$$\pm$$1.3$ \\
\cline{2-8}
\rowcolor{lightGray}
model & V
  & $2.2$$\pm$$2.9$ & $1.5$$\pm$$1.0$ & $1.6$$\pm$$0.9$ & $2.4$$\pm$$1.2$ & $2.0$$\pm$$0.6$ & $1.8$$\pm$$1.6$ \\
\cline{2-8}
\rowcolor{white}
 & F
  & $1.1$$\pm$$0.8$ & $1.2$$\pm$$0.8$ & $1.0$$\pm$$1.0$
  & N/A
  & $1.0$$\pm$$1.0$ & $1.1$$\pm$$0.9$ \\
\hline

\rowcolor{mediumGray}
GAN & A
  & $1.4$$\pm$$0.8$ & $1.8$$\pm$$1.0$ & $2.1$$\pm$$1.4$ & $2.5$$\pm$$2.4$ & $3.0$$\pm$$1.2$ & $2.0$$\pm$$1.5$ \\
\cline{2-8}
\rowcolor{lightGray}
model & V
  & $1.2$$\pm$$0.6$ & $1.8$$\pm$$1.9$ & $1.9$$\pm$$1.1$ & $2.6$$\pm$$2.2$ & $2.7$$\pm$$0.9$ & $1.9$$\pm$$1.6$ \\
\cline{2-8}
\rowcolor{white}
& F
  & $1.3$$\pm$$0.9$ & $1.0$$\pm$$0.6$ & $1.0$$\pm$$0.8$
  & N/A
  & $1.2$$\pm$$0.9$ & $1.1$$\pm$$0.8$ \\
\hline

\rowcolor{mediumGray}
Ensemble & A
  & $1.6$$\pm$$1.0$ & $1.5$$\pm$$0.7$ & $2.0$$\pm$$1.2$ & $2.6$$\pm$$2.5$ & $2.8$$\pm$$1.1$ & $1.9$$\pm$$1.5$ \\
\cline{2-8}
\rowcolor{lightGray}
prediction & V
  & $1.4$$\pm$$1.2$ & $1.5$$\pm$$1.5$ & $1.6$$\pm$$0.9$ & $2.7$$\pm$$2.1$ & $2.1$$\pm$$0.6$ & $1.7$$\pm$$1.5$ \\
\cline{2-8}
\rowcolor{white}
(3 models) & F
  & $0.8$$\pm$$0.9$
  & $1.0$$\pm$$0.5$
  & $0.8$$\pm$$0.7$
  & N/A
  & $0.8$$\pm$$0.7$ & $0.8$$\pm$$0.7$ \\
\hline

\rowcolor{mediumGray}
CBMT\cite{tang2023source} & A
  & $1.2$$\pm$$0.5$ & $2.0$$\pm$$0.8$ & $1.8$$\pm$$0.4$ & $1.8$$\pm$$0.6$ & $1.7$$\pm$$0.4$ & $1.7$$\pm$$0.6$ \\
\cline{2-8}
\rowcolor{lightGray}
w/ ensemble & V
  & $1.3$$\pm$$0.5$ & $1.8$$\pm$$0.7$ & $1.7$$\pm$$0.5$ & $2.1$$\pm$$0.8$ & $1.5$$\pm$$0.3$ & $1.7$$\pm$$0.6$ \\
\cline{2-8}
\rowcolor{white}
prediction & F
  & N/A & $1.4$$\pm$$0.6$ & $1.5$$\pm$$1.5$
  & N/A
  & N/A & $1.4$$\pm$$1.1$ \\
\hline

\rowcolor{mediumGray}
DPL\cite{chen2021source} w/ & A
  & $1.6$$\pm$$0.8$ & $2.2$$\pm$$1.2$ & $2.5$$\pm$$1.5$
  & $1.5$$\pm$$0.7$ & $3.2$$\pm$$1.1$ & $2.1$$\pm$$1.2$ \\
\cline{2-8}
\rowcolor{lightGray}
integrated & V
  & $1.1$$\pm$$0.6$ & $2.5$$\pm$$2.9$ & $2.1$$\pm$$1.2$
  & $1.7$$\pm$$0.8$ & $2.7$$\pm$$0.7$ & $1.9$$\pm$$1.6$ \\
\cline{2-8}
\rowcolor{white}
label & F
  & $1.2$$\pm$$0.6$ & $1.4$$\pm$$0.8$ & $1.0$$\pm$$0.7$
  & N/A
  & $1.0$$\pm$$0.8$ & $1.1$$\pm$$0.7$ \\
\hline
\hline
\rowcolor{mediumGray}
 & A
   & $1.2$$\pm$$0.6$\textcolor{green!70!black}{\scriptsize$\smallscript\Delta$.5}
   & $1.5$$\pm$$0.7$\textcolor{green!70!black}{\scriptsize$\smallscript\Delta$.2}
   & $1.7$$\pm$$0.9$\textcolor{green!70!black}{\scriptsize$\smallscript\Delta$.4}
   & $1.4$$\pm$$0.8$\textcolor{green!70!black}{\scriptsize$\smallscript\Delta$.5}
   & $2.3$$\pm$$0.8$\textcolor{green!70!black}{\scriptsize$\smallscript\Delta$.4}
   & $1.5$$\pm$$0.8$\textcolor{green!70!black}{\scriptsize$\smallscript\Delta$.4} \\
\cline{2-8}
\rowcolor{lightGray}
\textbf{GrInAdapt} & V
  & $1.1$$\pm$$0.7$\textcolor{green!70!black}{\scriptsize$\smallscript\Delta$.4}
  & $1.3$$\pm$$0.9$\textcolor{green!70!black}{\scriptsize$\smallscript\Delta$.7}
  & $1.5$$\pm$$0.8$\textcolor{green!70!black}{\scriptsize$\smallscript\Delta$.5}
  & $1.5$$\pm$$0.7$\textcolor{green!70!black}{\scriptsize$\smallscript\Delta$.6}
  & $1.9$$\pm$$0.7$\textcolor{green!70!black}{\scriptsize$\smallscript\Delta$.9}
  & $1.4$$\pm$$0.8$\textcolor{green!70!black}{\scriptsize$\smallscript\Delta$.5} \\
\cline{2-8}
\rowcolor{white}
\textbf{(Ours)} & F
  & $1.0$$\pm$$0.5$\textcolor{green!70!black}{\scriptsize$\smallscript\Delta$.8}
  & $1.0$$\pm$$0.4$\textcolor{green!70!black}{\scriptsize$\smallscript\Delta$.0}
  & $0.7$$\pm$$0.4$\textcolor{green!70!black}{\scriptsize$\smallscript\Delta$.2}
  & N/A
  & $0.7$$\pm$$0.4$\textcolor{green!70!black}{\scriptsize$\smallscript\Delta$.2}
  & $0.8$$\pm$$0.4$\textcolor{green!70!black}{\scriptsize$\smallscript\Delta$.4} \\
\hline

\end{tabular}
}
\end{table}

%% file: dice_assd_table_results/supp_both_site_assd_1f.tex
\begin{table}[t]
\setlength{\tabcolsep}{0.5pt}
\centering
\caption{Full results of cross-site methods comparison on the DSC score and ASSD distances A, V, F stand for artery, vein and FAZ. \textcolor{green!70!black}{$\Delta$} indicates improvements compared to the source model. The fomrat follows mean $\pm$ standard deviation.}
\label{tab:supp_both_only_site_assd_all_1deci}
\small
{\fontsize{8pt}{10pt}\selectfont
\begin{tabular}{c|c|ccc|ccc|c} 
\hline
\multirow{2}{*}{Methods} & \multirow{2}{*}{C}
  & \multicolumn{3}{c|}{\cite{tang2023source} DSC (\%) $\uparrow$}
  & \multicolumn{4}{c}{\cite{tang2023source} ASSD (pixel) $\downarrow$} \\ 
\cline{3-9}
& & UW & UAB & UCSD & UW & UAB & UCSD & All \\
\hline

\rowcolor{mediumGray}
Inegrated & A 
    & $70.2$$\pm$$7.8$ & $74.3$$\pm$$6.7$ & $65.7$$\pm$$17.8$
    & $1.4$$\pm$$0.7$ & $1.1$$\pm$$0.5$ & $1.9$$\pm$$1.8$ & $1.5$$\pm$$1.2$ \\
\cline{2-9}
\rowcolor{lightGray}
Label & V
    & $70.7$$\pm$$7.6$ & $78.1$$\pm$$6.1$ & $69.6$$\pm$$17.8$
    & $1.3$$\pm$$0.6$ & $1.0$$\pm$$0.5$ & $1.7$$\pm$$1.7$ & $1.3$$\pm$$1.1$ \\
\cline{2-9}
\rowcolor{white}
(Fig. 1d)  & F
    & $73.9$$\pm$$33.8$ & $84.8$$\pm$$14.0$ & $88.4$$\pm$$5.4$
    & $1.2$$\pm$$0.7$ & $1.7$$\pm$$2.5$ & $1.3$$\pm$$0.6$ & $1.4$$\pm$$1.5$ \\
\hline
\hline

\rowcolor{mediumGray}
Source & A
    & $67.0$$\pm$$7.8$ & $70.0$$\pm$$7.2$ & $63.8$$\pm$$7.5$
    & $2.0$$\pm$$1.2$ & $1.5$$\pm$$0.7$ & $2.2$$\pm$$1.3$ & $1.9$$\pm$$1.1$ \\
\cline{2-9}
\rowcolor{lightGray}
Model & V
    & $66.7$$\pm$$7.2$ & $72.3$$\pm$$6.7$ & $66.6$$\pm$$8.5$
    & $1.9$$\pm$$1.1$ & $1.6$$\pm$$1.0$ & $2.3$$\pm$$1.8$ & $1.9$$\pm$$1.3$ \\
\cline{2-9}
\rowcolor{white}
($f_{\theta_S}$) & F
    & $74.1$$\pm$$33.4$ & $90.6$$\pm$$4.9$ & $89.9$$\pm$$7.2$
    & $1.5$$\pm$$2.4$ & $0.9$$\pm$$0.7$ & $1.0$$\pm$$0.7$ & $1.2$$\pm$$1.6$ \\
\hline

\rowcolor{mediumGray}
NoNorm & A
    & $65.4$$\pm$$8.7$ & $69.7$$\pm$$6.9$ & $63.0$$\pm$$6.8$
    & $2.3$$\pm$$1.6$ & $1.6$$\pm$$0.7$ & $2.5$$\pm$$1.2$ & $2.2$$\pm$$1.3$ \\
\cline{2-9}
\rowcolor{lightGray}
Model & V
    & $66.5$$\pm$$7.4$ & $72.0$$\pm$$5.9$ & $66.2$$\pm$$9.3$
    & $1.7$$\pm$$0.8$ & $1.4$$\pm$$1.0$ & $2.4$$\pm$$2.5$ & $1.8$$\pm$$1.6$ \\
\cline{2-9}
\rowcolor{white}
& F
    & $81.3$$\pm$$26.3$ & $90.7$$\pm$$6.0$ & $86.8$$\pm$$9.8$
    & $1.0$$\pm$$0.6$ & $0.9$$\pm$$0.9$ & $1.4$$\pm$$1.1$ & $1.1$$\pm$$0.9$ \\
\hline

\rowcolor{mediumGray}
GAN & A
    & $66.8$$\pm$$8.8$ & $70.7$$\pm$$7.5$ & $59.3$$\pm$$16.7$
    & $2.0$$\pm$$1.2$ & $1.4$$\pm$$0.7$ & $2.6$$\pm$$2.1$ & $2.0$$\pm$$1.5$ \\
\cline{2-9}
\rowcolor{lightGray}
Model & V
    & $67.0$$\pm$$7.8$ & $72.8$$\pm$$6.7$ & $62.8$$\pm$$16.9$
    & $1.8$$\pm$$1.1$ & $1.6$$\pm$$1.5$ & $2.3$$\pm$$2.0$ & $1.9$$\pm$$1.6$ \\
\cline{2-9}
\rowcolor{white}
 & F
    & $74.9$$\pm$$31.7$ & $90.3$$\pm$$5.5$ & $88.4$$\pm$$9.1$
    & $1.2$$\pm$$1.0$ & $0.9$$\pm$$0.5$ & $1.1$$\pm$$0.7$ & $1.1$$\pm$$0.8$ \\
\hline

\rowcolor{mediumGray}
Ensemble & A
    & $68.1$$\pm$$8.6$ & $72.4$$\pm$$7.3$ & $61.1$$\pm$$17.4$
    & $1.9$$\pm$$1.2$ & $1.4$$\pm$$0.7$ & $2.5$$\pm$$2.1$ & $1.9$$\pm$$1.5$ \\
\cline{2-9}
\rowcolor{lightGray}
prediction & V
    & $68.5$$\pm$$7.4$ & $74.8$$\pm$$6.4$ & $64.8$$\pm$$17.5$
    & $1.6$$\pm$$0.9$ & $1.3$$\pm$$0.9$ & $2.3$$\pm$$2.1$ & $1.7$$\pm$$1.5$ \\
\cline{2-9}
\rowcolor{white}
(3 models) & F
    & $80.0$$\pm$$28.2$ & $93.4$$\pm$$3.8$ & $91.7$$\pm$$7.4$
    & $1.0$$\pm$$0.8$ & $0.7$$\pm$$0.6$ & $0.9$$\pm$$0.7$ & $0.8$$\pm$$0.7$ \\
\hline

\rowcolor{mediumGray}
CBMT\cite{tang2023source} & A
    & $67.0$$\pm$$5.7$ & $68.6$$\pm$$5.3$ & $65.2$$\pm$$6.0$
    & $1.8$$\pm$$0.5$ & $1.4$$\pm$$0.3$ & $2.0$$\pm$$0.8$ & $1.7$$\pm$$0.6$ \\
\cline{2-9}
\rowcolor{lightGray}
w/ ensemble & V
    & $67.4$$\pm$$5.0$ & $70.4$$\pm$$4.8$ & $66.5$$\pm$$6.4$
    & $1.7$$\pm$$0.5$ & $1.5$$\pm$$0.4$ & $1.9$$\pm$$0.9$ & $1.7$$\pm$$0.6$ \\
\cline{2-9}
\rowcolor{white}
prediction & F
    & $40.5$$\pm$$46.7$ & $45.6$$\pm$$45.0$ & $43.5$$\pm$$45.0$
    & $1.0$$\pm$$0.5$ & $1.7$$\pm$$1.8$ & $1.6$$\pm$$0.7$ & $1.4$$\pm$$1.1$ \\
\hline

\rowcolor{mediumGray}
DPL\cite{chen2021source} w/ & A
    & $69.6$$\pm$$8.4$ & $72.4$$\pm$$8.0$ & $68.6$$\pm$$6.1$
    & $2.1$$\pm$$1.2$ & $1.8$$\pm$$1.2$ & $2.3$$\pm$$1.2$ & $2.1$$\pm$$1.2$ \\
\cline{2-9}
\rowcolor{lightGray}
integrated & V
    & $69.3$$\pm$$7.5$ & $74.1$$\pm$$7.3$ & $69.9$$\pm$$8.0$
    & $1.8$$\pm$$1.0$ & $1.7$$\pm$$1.8$ & $2.2$$\pm$$2.1$ & $1.9$$\pm$$1.6$ \\
\cline{2-9}
\rowcolor{white}
label & F
    & $81.5$$\pm$$20.3$ & $88.5$$\pm$$6.2$ & $88.5$$\pm$$9.1$
    & $1.1$$\pm$$0.5$ & $1.1$$\pm$$0.7$ & $1.2$$\pm$$0.9$ & $1.1$$\pm$$0.7$ \\
\hline

\rowcolor{mediumGray}
\textbf{GrIn}& A
    & \textbf{$70.3$$\pm $$7.3$}\textcolor{green!70!black}{\scriptsize$\smallscript\Delta$3.3}
      & \textbf{$73.3$$\pm$$6.7$}\textcolor{green!70!black}{\scriptsize$\smallscript\Delta$3.3}
      & \textbf{$68.1$$\pm$$6.2$}\textcolor{green!70!black}{\scriptsize$\smallscript\Delta$4.3}
    & \textbf{$1.6$$\pm$$0.8$}\textcolor{green!70!black}{\scriptsize$\smallscript\Delta$.4}
      & \textbf{$1.2$$\pm$$0.5$}\textcolor{green!70!black}{\scriptsize$\smallscript\Delta$.3}
      & \textbf{$1.7$$\pm$$0.9$}\textcolor{green!70!black}{\scriptsize$\smallscript\Delta$.5}
      & \textbf{$1.5$$\pm$$0.8$}\textcolor{green!70!black}{\scriptsize$\smallscript\Delta$.4} \\
\cline{2-9}
\rowcolor{lightGray}
\textbf{Adapt} & V
    & \textbf{$70.3$$\pm$$6.6$}\textcolor{green!70!black}{\scriptsize$\smallscript\Delta$3.6}
      & \textbf{$75.1$$\pm$$5.9$}\textcolor{green!70!black}{\scriptsize$\smallscript\Delta$2.8}
      & \textbf{$70.2$$\pm$$7.1$}\textcolor{green!70!black}{\scriptsize$\smallscript\Delta$3.6}
    & \textbf{$1.4$$\pm$$0.7$}\textcolor{green!70!black}{\scriptsize$\smallscript\Delta$.5}
      & \textbf{$1.1$$\pm$$0.6$}\textcolor{green!70!black}{\scriptsize$\smallscript\Delta$.5}
      & \textbf{$1.6$$\pm$$0.9$}\textcolor{green!70!black}{\scriptsize$\smallscript\Delta$.7}
      & \textbf{$1.4$$\pm$$0.8$}\textcolor{green!70!black}{\scriptsize$\smallscript\Delta$.5} \\
\cline{2-9}
\rowcolor{white}
\textbf{(Ours)} & F
    & \textbf{$84.7$$\pm$$20.$}\textcolor{green!70!black}{\scriptsize$\smallscript\Delta$10.6}
      & \textbf{$92.2$$\pm$$3.5$}\textcolor{green!70!black}{\scriptsize$\smallscript\Delta$1.6}
      & \textbf{$91.5$$\pm$$5.0$}\textcolor{green!70!black}{\scriptsize$\smallscript\Delta$1.6}
    & \textbf{$0.9$$\pm$$0.4$}\textcolor{green!70!black}{\scriptsize$\smallscript\Delta$.6}
      & \textbf{$0.7$$\pm$$0.3$}\textcolor{green!70!black}{\scriptsize$\smallscript\Delta$.2}
      & \textbf{$0.9$$\pm$$0.5$}\textcolor{green!70!black}{\scriptsize$\smallscript\Delta$.1}
      & \textbf{$0.8$$\pm$$0.4$}\textcolor{green!70!black}{\scriptsize$\smallscript\Delta$.4} \\
\hline

\end{tabular}
}
\end{table}

%% file: Supp/Supp_reg.tex
\section{Registration Algorithm Used in GrInAdapt}
\label{sec:supp_reg}
In this section, we provide more details of the proposed registration process of our GrInAdapt framework. Specifically, we present specifics of the homography decomposition and validation, as well as the multi-trial registration with adaptive anchor selection.

\subsection{Registration Algorithm, Homography Decomposition and Validation}
The registration process employed an affine transformation model to align images from different domains. For each pair of images, we estimated a homography matrix $H \in \mathbb{R}^{3 \times 3}$ using AKAZE keypoint detection and RANSAC-based matching. To ensure anatomical plausibility of the transformation, we decomposed $H$ into interpretable components and applied specific validation thresholds.
Given a homography matrix generated based on the set of the matched points:
\begin{equation}
H = \begin{bmatrix} 
h_{11} & h_{12} & h_{13} \\ 
h_{21} & h_{22} & h_{23} \\ 
h_{31} & h_{32} & h_{33} 
\end{bmatrix}
\end{equation}
We first normalized $H$ such that $h_{33} = 1$. The decomposition extracted the following components:
\begin{itemize}
    \item \textbf{Translation vector:} $\mathbf{t} = (t_x, t_y)$ where $t_x = h_{13}$ and $t_y = h_{23}$.
    
    \item \textbf{Linear transformation matrix:} $A = \begin{bmatrix} h_{11} & h_{12} \\ h_{21} & h_{22} \end{bmatrix}$.
    
    \item \textbf{Scale factors:} $s_x = \|A_{:,1}\|_2$ and $s_y = \|A_{:,2}\|_2$, representing the L2 norms of the first and second columns of $A$.
    
    \item \textbf{Rotation:} $\theta = \arctan2(R_{21}, R_{11}) \cdot \frac{180}{\pi}$ where $R = A \cdot \text{diag}(1/s_x, 1/s_y)$ is the normalized transformation matrix.
    
    \item \textbf{Shear factor:} $\gamma = \frac{A_{11}A_{12} + A_{21}A_{22}}{s_x s_y}$, quantifying the non-orthogonality of the transformation.
    
    \item \textbf{Perspective distortion:} $p = \sqrt{h_{31}^2 + h_{32}^2}$, measuring the deviation from the affinity.
\end{itemize}

A homography is considered valid if all components satisfy anatomically plausible thresholds. Let $\alpha$, $\beta$, $\tau$, $\phi$, and $\rho$ be different thresholds corresponding to different decompositions. The scale factors must be within a reasonable range ($\alpha_1 \leq s_x, s_y \leq \alpha_2$), limiting excessive expansion or compression. The rotation angle must be modest ($|\theta| \leq \beta$), preventing unrealistic rotations. The translation components must be bounded ($|t_x|, |t_y| \leq \tau$), ensuring that the transformation does not alter the structures excessively. The shear must be limited ($|\gamma| \leq \phi$), avoiding non-rigid deformations that could distort vessel morphology. Finally, the perspective distortion must be minimal ($p \leq \rho$), maintaining the approximately affine nature of retinal transformations.

We set $\alpha_1 = 0.5$, $\alpha_2 = 2.0$, $\beta = 15^{\circ}$, $\phi = 0.5$, and $\rho = 0.01$. We didn't restrict $\tau$ in this study ro provide more flexibility to our registration algorithm. These thresholds were empirically determined to accommodate the expected range of anatomical variations while preventing implausible transformations.

\subsection{Multi-trial Registration with Adaptive Anchor Selection}
To enhance registration robustness, we implemented a multi-trial approach with adaptive anchor selection. Given a subject $s_k$ with $N_{s_k}$ images, we initially chose one image as the anchor and attempted to register all other images to it. The registration produced a set of homography matrices $\{H_j\}_{j=1}^{N_{s_k}-1}$.

If the initial registration validation failed, we adaptively selected a new anchor based on the translation clustering analysis. Let $\mathbf{t} = \{(t_x^j, t_y^j)\}_{j=1}^{N_{s_k}-1}$ be the set of translation vectors extracted from the homographies. 
We performed a translation clustering analysis using the K-means algorithm and picked the centroid $\mu_t$ with the largest number of samples. 
We then computed the distances between each translation vector and the centroid $d_j = \|(t_x^j, t_y^j) - \mu_t\|_2$.
Next, we selected the image corresponding to the smallest distance as the new anchor $j^{\text{new}} = \arg\min_{j \in \{1, \ldots, N_{s_k}-1\} \setminus \mathcal{U}} d_j$, where $\mathcal{U}$ is the set of previously used anchor indices. This process continued until either a valid registration was found or all possible anchors were been evaluated. If all anchors led to failed registration, this subject was considered as failed and was not be included in the study.

On the other hand, for subjects with successful registration passing the validation process, we optionally performed an additional registration trial with a different anchor image, typically the one with the second-smallest distance in the first trial. This extra step generated one more valid registration configuration. We selected the configuration with the smaller cluster center distance. In practice, we found this dual-trial approach significantly improved the registration quality by identifying the anchor that provides the most stable transformation across all images. If there was only one successful configuration, we directly picked it. The selected configuration was then used for label integration and model adaptation.


%% file: Supp/Supp_integrate.tex
\section{More Details of the Integration Process in GrInAdapt}
In the supplementary material, we provide a comprehensive description of the integration process. Specifically, we describe our region level strategy for macula, optic disc, and other region by doing two registrations covering different images using the algorithm described in section \ref{sec:supp_reg}.

\noindent\textbf{Macular Region.}
We first registered the four OCTA macula-centered images (three 6mm $\times$ 6mm and one 12mm $\times$ 12mm). Each 6mm $\times$ 6mm probability map is evenly padded with zeroes to a shape of (512, 512), while the 12mm $\times$ 12mm OCTA macula probability map was interpolated up to a shape of (512, 512). After registering to the same space, we cropped the central $(256, 256)$ region from each registered image to extract the macula region. For pixels where at least one domain had the prediction of Artery, Vein, or FAZ, we averaged the corresponding probability vectors. The background and capillary class are excluded because of their high variability due to device-specific noise. 

\noindent\textbf{Optic disc region and other region.}
Outside the macula region, we performed a second registration for the larger region. Specifically, we used the combined macula probability map, optic disc centered OCTA, 12mm$\times$12mm OCTA, and CFP vessel map prediction.

After the registration, we first generated a mask by selecting pixels in that do not overlap with the macula region. Only shared classes - artery and vein - were merged together. For the optic disc region, the prediction of optic disc centered OCTA and CFP prediction were merged. For the other region, the 12mm$\times$12mm OCTA scans and CFP prediction were merged.

\noindent\textbf{Final label generation and reversion.}
We combined probability maps for all three regions to get the fully-merged probability map. We next transformed the label back based on the inverse homography via Eqn. (\ref{eqn:merge}). For macula region, the homography was the one generated by macula-only registration. For other two regions, the homography was the one generated throughout the second full registration.

%% file: Supp/Supp_adapt.tex
\section{More details of the Adaptation Process in GrInAdapt}

In this section, we provide more details of the proposed adaptation process of our GrInAdapt framework. Specifically, we present the details of the data augmentation techniques, the modified pseudo-label generation process for our teacher model in order to accommodate the usage of \cite{chen2021source} and \cite{tang2023source} as suitable baselines. We also present other undescribed details of our teacher-student model implementation.

\subsection{Weak-Strong Data Augmentation Strategy for the OCTA and OCT input}
Our teacher-student framework is inspired by \cite{tang2023source}. However, the weak and strong augmentation used in \cite{tang2023source} are not directly applicable to 3D OCTA flow and 3D structural OCT volumes.
To make sure the weak-strong augmentation can be used in our framework, and also to enhance the robustness of the adaptation process, we employed distinct augmentation strategies for strong augmentation during the teacher-student learning process.

\noindent\textbf{Gaussian noise.}
For both 3D OCTA flow, 3D structural OCT volumes and their corresponding 2D projection maps, we first applied a Gaussian noise with a variance of $\sigma^2$. Let $\epsilon \sim \mathcal{N}(0, \sigma^2$) be the noise sampled from the Gaussian distribution, we directly let $x' = x + \epsilon$ be the augmented volumes and images.
The weakly augmented teacher model has a $\sigma_t$ of 0.01, while the strongly augmented student model used a schedule of $\sigma_s$:   
\begin{equation}
\sigma_s = 
\begin{cases}
0.2, & \text{with probability } 0.5, \\
0.1, & \text{with probability } 0.5.
\end{cases}
\end{equation}

\noindent\textbf{Additional Augmentation for 2D Projection Maps.}
For the 2D projection maps, we implemented an additional contrast adjustment with a light adjustment factor $a = 0.2$ and a random contrast factor $C = 1 + 2 \cdot a \cdot \left(u - \frac{1}{2}\right)$ derived from an uniform distribution $u \sim \mathcal{U}(0,1)$. The contrast-adjusted \textit{en face} images were then computed as $x'' = x' \cdot C$.

\subsection{Pseudo-Label Generation Process for the Teacher Model}

Following \cite{tang2023source}, the teacher model generated pseudo-labels that serve as targets for the student model. Still, the design used in \cite{tang2023source} was for optic disc and cup adaptation, so it can not be used in our task. We thus employed a class-specific thresholding strategy to improve the quality of these pseudo-labels. Specifically, our thresholding strategy was carefully designed to account for the unique characteristics of each class:
\begin{itemize}
\item \textbf{Background (Class 0) and Capillary (Class 1):} We applied no threshold for these classes since they constitute the majority of pixels in retinal images and exhibit higher prediction consistency. Given the inherent uncertainty in model predictions for these widespread structures, we opted to preserve all predictions rather than risk information loss through aggressive thresholding. Still, ambiguous pixels will be assigned to background class, which is further discussed below.

\item \textbf{Artery (Class 2) and Vein (Class 3):} We implemented a fixed threshold of $\tau_{A} = \tau_V = 0.5$, which represents a 2.5× confidence level over the random chance baseline for our 5-class CAVF task. This ensures that only confident vessel predictions participate in the adaptation process, reducing the risk of error propagation in these critical anatomical structures.

\item \textbf{Foveal Avascular Zone (Class 4):} Given that the FAZ is anatomically constrained to the central retinal region for macula-centered OCTA scans, for pixel $p$, we employed a spatially-aware, region-adaptive thresholding approach based on the Euclidean distance $d(p)$ from 
$p$ to the image center $(\frac{H}{2}, \frac{W}{2})$. Let $d_c=\sqrt{\frac{H^2+W^2}{4}}$, we have:
\begin{equation}
\tau_{F}(p) =
\begin{cases}
0.2, & \text{if } d(p) < 0.4 \times d_c, \\
0.7, & \text{if } 0.4 \times d_c \leq d(p) < 0.5 \times d_c, \\
0.95, & \text{if } d(p) \geq 0.5 \times d_c.
\end{cases}
\end{equation}
\end{itemize}
This stratified threshold accounts for the high likelihood of FAZ in the central region (requiring only 0.2 confidence, equivalent to $1\times$ confidence level), with progressively stricter thresholds (0.7 and 0.95, $3.5\times$ and $4.75\times$ confidence level) as distance increases, effectively eliminating false positives in peripheral regions where FAZ should not appear. For the FAZ label in the Optic disc centered scans, we removed all of them because no FAZ area would be presented.

\noindent\textbf{Ambiguity Resolution.}
If no class meets its respective threshold condition at a pixel $p$ (i.e., the pixel prediction is "ambiguous"), the pixel would be assigned to the background class (Class 0) by default. This conservative approach is simlar to \cite{tang2023source}, preventing the propagation of uncertain predictions during adaptation.

\subsection{Teacher-Student Model Implementation}
We combined two different weighting strategies for the loss function. The first strategy directly followed \cite{tang2023source}, where we used the pixel-level class-balanced loss weight design by considering the number of pixels in each class. This addresses the class imbalance among our five classes. We further introduced a second holistic class-specific reweighting design for the Artery, Vein and FAZ class (Class 2,3,4). 
Specifically, we utilized two complementary weight vectors: the teacher model's predicted pseudo-label class weights $[1, 1, w^t_{a}, w^t_{v}, w^t_{f}]$ and the integrated label class weights $[1, 1, w^{int}_{a}, w^{int}_{v}, w^{int}_{f}]$. For each class used this schedule, let $w^t_{c,\min}$ and $w^t_{c,\max}$ be the minimum and maximum value of the weight of the teacher model prediction, $w^{int}_{c,\min}$ and $w^{int}_{c,\max}$ be the minimum and maximum value of the weight of integrated labels. The weights followed a cosine-based scheduling defined by:
\begin{align}
w^t_{c} &= w^{t}_{c,\min} + (w^{t}_{c,\max} - w^{t}_{c,\min}) \cdot \eta, c \in \{\text{Artery}, \text{Vein}, \text{FAZ}\} \\
w^{int}_{c} &= w^{int}_{c,\max} - (w^{int}_{c,\max} - w^{int}_{c,\min}) \cdot \eta, c \in \{\text{Artery}, \text{Vein}, \text{FAZ}\}
\end{align}
where $\eta$ is a cosine annealing schedule from 0 to 1 across $E$ epochs.
 For Artery and Vein, we set $w^{t}_{c,\min} = w^{int}_{c,\min} = 0.5$, $w^{t}_{c,\max} = w^{int}_{c,\max} = 1.5$. For FAZ, we set $w^{t}_{\min} = 0$, $w^{t}_{c,\max} = 1.2$, $w^{int}_{c,\min} = 0.8$, and $w^{int}_{c,\max} = 2$.
This dynamic scheduling strategy enables the model to initially rely more heavily on the integrated labels (which provide superior Artery, Vein and FAZ delineation) while gradually transitioning to the teacher model's predictions as training progresses.

%% file: Supp/Supp_other.tex
\section{Auxiliary Details of CFP Modality}
\noindent\textbf{Models $f_{\theta_A}$.} The CFP model architecture used was the GAN model from \cite{zhou2021learning}. We directly used the three pre-trained model weights trained on three different datasets: the DRIVE-AV dataset \cite{hu2013automated}, the LES-AV \cite{orlando2018towards} dataset, and the HRF-AV \cite{odstrcilik2013retinal} dataset. We used average Artery-Vein predictions of three models for each CFP image.

\noindent\textbf{Paired CFP images in AI-READI dataset.}
In addition to OCTA volumes, AI-READI also contains 4 types of CFP images: Maestro2 Macula, Maestro2 Wide Field, Triton Macula 12x12, and Triton Macula 6x6 images. For each subject, we used the average prediction of these four by first doing a registration on them. In summary, each of the labels used for integration can be treated as an average over 12 predictions. 